\documentclass[aps,prd,showpacs]{revtex4}

\usepackage[small]{caption}
\usepackage{amsmath,amsfonts,amscd,amssymb,epsf, epsfig,mathrsfs,graphicx}

\newcommand{\pa}{\partial}

\newcommand{\vep}{\varepsilon}

\begin{document}

\title{ Casimir interaction between  a cylinder and a plate at finite temperature: Exact results and comparison to proximity force approximation}

 \author{L. P. Teo}\email{ LeePeng.Teo@nottingham.edu.my}
\address{Department of Applied Mathematics, Faculty of Engineering, University of Nottingham Malaysia Campus, Jalan Broga, 43500, Semenyih, Selangor Darul Ehsan, Malysia.}
 \begin{abstract}
 We study the finite temperature Casimir interaction between a cylinder and a plate using the exact formula derived from the Matsubara representation and the functional determinant representation. We consider the scalar field with Dirichlet and Neumann boundary conditions.  The asymptotic expansions of the Casimir energy and the Casimir force when the separation $a$ between the cylinder and the plate is small are derived. As in the zero temperature case, it is found that  the leading terms of the Casimir energy and the Casimir force agree with those derived from the proximity force approximation when     $rT\gg 1$, where $r$ is the radius of the cylinder. When $aT\ll 1\ll rT$ (the medium temperature region), the leading term of the Casimir energy is of order $T^{\frac{5}{2}}$ whereas for the Casimir force, it is of order $T^{\frac{7}{2}}$. In this case, the leading terms are independent of the separation $a$. When $1\ll aT\ll rT$ (the high temperature region), the dominating terms of the Casimir energy and the Casimir force come from the zeroth Matsubara frequency. In this case, the leading terms are linear in $T$, but for the energy, it is inversely proportional to $a^{\frac{3}{2}}$, whereas for the force, it is inversely proportional to   $a^{\frac{5}{2}}$. The first order corrections to the proximity force approximations in different temperature regions are computed using perturbation approach. In the zero temperature case, the results agree with those derived in [Bordag, Phys. Rev. D \textbf{73}, 125018 (2006)].
 \end{abstract}

\maketitle
\section{Introduction }

In the past ten year, there has been a tremendous interest in the Casimir interaction between two objects. This is in part due to the advent in the  precise measurement of the Casimir force \cite{48,34,35,36,37,38,39,40,42}. The  breakthrough in the methods for effective computations of the Casimir force has helped to make possible the exact and numerical computations of the Casimir force beyond the precision afforded by the proximity force approximation. Some of the  methods developed for  computing the Casimir interactions
include the semi-classical approximation \cite{43,44}, the optical path method \cite{45,46,47}, the worldline approach \cite{14,15,16,17}, the functional determinant or the multiple scattering method \cite{1,27,3,9,10,11,12,32,13}, and the exact mode summation method \cite{21,2}.  These methods have been used to compute the Casimir interactions of a number of geometric configurations. Among the most popularly studied configurations are  the sphere-plane \cite{43,44,45,46,47,1,14,15,17,13,27,19,23,20,29,31},  the cylinder-plane \cite{15,17,1,3,13,21,22,2,4},  the cylinder-cylinder \cite{21,2,22,13,30,32,33} and the sphere-sphere \cite{27,32,44,9,10,11,12,24,25} configurations.  It has been shown that all the different approaches give  leading order approximations to the Casimir force at zero temperature that agree  with those obtained using proximity force approximations. There is an intense interest in computing the corrections to the proximity force approximations.

Among the methods mentioned above,   the functional determinant or multiple scattering approach and the mode summation approach can give  an exact formula for the Casimir energy or the Casimir force that is valid at all separations between the objects. In general, it is easy to deduce the asymptotic behavior of the Casimir force from the exact formula when the separation between the objects are large. When the separation is small, this become numerically complicated. Therefore, this  hampers the comparison between the results obtained numerically using approximation methods and the   result computed from the exact formula. In the pioneering work \cite{1}, Bordag has introduced an analytical method to obtain the first order correction to the proximity force approximation for the configuration of a cylinder in front of a plane. This is later generalized to the configuration of a sphere in front of a plane \cite{23,31}.

In recent years, there is a growing interest in the thermal correction to the Casimir interaction from both the theoretical and the experimental sides \cite{48,39}. However, the finite temperature Casimir interaction between two compact objects have been less studied compared to the zero temperature case. Weber and Gies \cite{7,6} extended the worldline algorithm to the finite temperature case and showed that there is an interesting interplay between temperature and geometry. Both the sphere-plane and the cylinder-plane configurations were considered. However, as in the zero temperature case, so far the worldline method has only been developed for the case of Drichlet boundary conditions. In \cite{49,28}, the functional determinant representation of the finite temperature Casimir interaction between a sphere and a plane was derived which also took into account dielectric materials. The Casimir interactions at medium and large separations were obtained half analytically and half numerically.
In \cite{5,26}, Bordag and  Pirozhenko studied analytically the finite temperature Casimir interaction between a sphere and a plane. The asymptotic behaviors at small separation is computed to the leading orders and they are found to be coincide with the proximity force approximations at medium and high temperature \cite{5}.

Among the different two-object configurations, the sphere-plane configuration is the most studied since this has been extensively used in Casimir experiments \cite{48,39}. However, it has been proposed that the cylinder-plane setup has some of the advantages of both the parallel-plane and the sphere-plane setups \cite{50,41,42}. In \cite{18}, Klimchitskaya and Romero have explored the possibility of obtaining stronger constraints on Yukawa-type corrections to Newtonian gravity from measuring the Casimir force between a cylinder and a plane. Thus the cylinder-plane configuration is becoming increasingly important in the study of Casimir effect. The exact Casimir interaction between a cylinder and a plane at finite temperature has been discussed briefly in \cite{3}. The low temperature asymptotic behavior of the thermal correction has been obtained for large separation. So far no work has considered the exact finite temperature Casimir interaction between a cylinder and a plane at small separation. The purpose of this paper is to fill in this gap.

In this paper, we derive the functional determinant representation of the finite temperature Casimir interaction between a cylinder and a plane using the Matsubara formalism. The exact expression allows us to conclude that the Casimir force is attractive at any temperature for both the Dirichlet and the Neumann boundary conditions, and hence for the perfect electric conductor condition. We then derive the small separation asymptotic behaviors of the Casimir force in different temperature regions. The leading order terms are compared to those obtained using proximity force approximation. The first order corrections are also computed analytically.

In this paper, we use units with $\hbar=c=k_B=1$.

 \section{Proximity force approximation}
 In this section, we use the proximity force approximation (PFA) to find the leading asymptotic behavior of the Casimir force between a cylinder and a plate when the separation between them is small. This has been considered in \cite{6} but we use a different approach here that would show explicitly the leading contributions in each of the temperature regions.

  \begin{figure}[h]
\epsfxsize=0.5\linewidth \epsffile{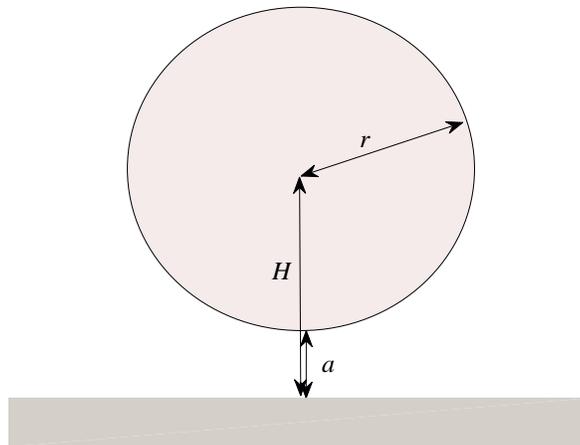} \caption{\label{f1} The cross section of a cylinder in front of a plate. }\end{figure}

As shown in Fig. \ref{f1}, a cylinder length $L$ and radius $r$ is placed in front of a plate. The distance between the cylinder and the plate is $a$, and the distance from the center of the cylinder to the plate is equal to $a+r$, which we denote by $H$.

The finite temperature Casimir force density for a pair of parallel Dirichlet/Neumann plates separated by a distance $d$ is given by \cite{8}:
\begin{equation}\label{eq4_6_1}\mathcal{F}_{\text{Cas}}^{\parallel}(d)=-\frac{\pi^2   }{480 d^4} -\frac{\pi^2 T^4}{90}+\frac{\pi T}{2d^3}\sum_{k=1}^{\infty}\sum_{l=1}^{\infty}\frac{k^2}{l}\exp\left(-\frac{\pi kl}{dT}\right),\end{equation}
or
\begin{equation}\label{eq4_6_2}\mathcal{F}_{\text{Cas}}^{\parallel}(d)=-\frac{\zeta_R(3) T  }{8\pi  d^3}-\frac{T}{\pi}\sum_{k=1}^{\infty}\sum_{l=1}^{\infty}\left(
\frac{2\pi^2l^2T^2}{kd}+\frac{ \pi l T}{k^2d^2}+\frac{1}{4k^3d^3}\right) e^{-4\pi kl dT}.\end{equation}
Eq. \eqref{eq4_6_1} is the low temperature expansion. It shows that in the low temperature region (i.e., $dT\ll 1$), the Casimir force density is dominated by the zero temperature term
\begin{equation}\label{eq4_7_1}
\mathcal{F}_{\text{Cas}}^{\parallel,T=0}(d)=-\frac{\pi^2   }{480 d^4}.
\end{equation}The leading term of the thermal correction is the $d$-independent term
\begin{equation}\label{eq4_7_2}\Delta_T\mathcal{F}_{\text{Cas}}^{\parallel } \sim -\frac{\pi^2 T^4}{90},\end{equation}
and the remaining terms go to zero exponentially fast as $dT\rightarrow0$. Eq. \eqref{eq4_6_2} is the high temperature expansion. It shows that in the high temperature region (i.e., $dT\gg 1$),
the Casimir force is dominated by the term
\begin{equation}\label{eq4_7_3}
\mathcal{F}_{\text{Cas}}^{\parallel,\text{cl}}(d)=-\frac{\zeta_R(3) T  }{8\pi  d^3},\end{equation}which is called the classical term. The remaining terms go to zero exponentially fast when $dT\rightarrow\infty$.

  \begin{figure}[h]
\epsfxsize=0.5\linewidth \epsffile{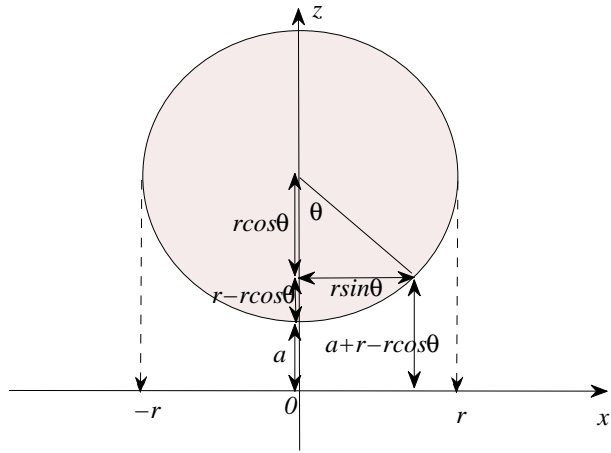} \caption{\label{f2} The cross section of the cylinder-plate system in the $xz$--plane.}\end{figure}
Applying the proximity force approximation to the cylinder-plate configuration with either Dirichlet or Neumann boundary conditions, we need to integrate the Casimir force density between the parallel plates $\mathcal{F}_{\text{Cas}}^{\parallel}(a)$ over the rectangle $\{(x,y)\;:\; -r\leq x\leq r,\;0\leq y\leq L\}$. At the point $(x,y)$, the distance between the cylinder and the plate is $a+r(1-\cos\theta)$, where $x=r\sin\theta$ (see Fig. \ref{f2}). Therefore, the proximity force approximation for the force is
\begin{equation*}
F_{\text{Cas}}^{ \text{PFA}}=  2L\int_{0}^r \mathcal{F}_{\text{Cas}}^{\parallel }(a+r(1-\cos\theta)) dx.
\end{equation*}
Let
$t=1-\cos\theta$.
Then
\begin{equation}\label{eq4_6_3}
\begin{split}
F_{\text{Cas}}^{ \text{PFA}}=  2Lr\int_0^1\mathcal{F}_{\text{Cas}}^{\parallel }(a+rt)(1-t)\frac{dt}{\sqrt{2t-t^2}}.
\end{split}
\end{equation}
For the Casimir energy, the proximity force approximation gives
\begin{equation*}
\begin{split}
E_{\text{Cas}}^{ \text{PFA}}=  2Lr\int_0^1\mathcal{E}_{\text{Cas}}^{\parallel }(a+rt)(1-t)\frac{dt}{\sqrt{2t-t^2}},
\end{split}
\end{equation*}where $\mathcal{E}_{\text{Cas}}^{\parallel }$ is the finite temperature Casimir energy density for a pair of parallel Dirichlet/Neumann plates given by
\begin{equation*}
\begin{split}
\mathcal{E}_{\text{Cas}}^{\parallel}(d)=& -\frac{\pi^2}{1440d^3}+\frac{\pi^2dT^4}{90}- \sum_{l=1}^{\infty}\sum_{k=1}^{\infty}\left(\frac{kT^2}{2l^2d}+\frac{T^3}{2\pi l^3}\right)e^{-\frac{\pi k l }{dT}},
\end{split}
\end{equation*}or
\begin{equation*}
\begin{split}
\mathcal{E}_{\text{Cas}}^{\parallel}(d)
 =&-\frac{T}{16\pi d^2}\zeta_R(3)-\frac{T}{\pi}\sum_{l=1}^{\infty}\sum_{k=1}^{\infty}\left(\frac{ \pi l T}{2 k^2d}+\frac{1}{8k^3d^2}\right)e^{-4\pi kl dT}.
\end{split}
\end{equation*}
Obviously,
$$F_{\text{Cas}}^{ \text{PFA}}=-\frac{\pa E_{\text{Cas}}^{ \text{PFA}}}{\pa a}.$$
As in \cite{5}, we study the proximity force approximation in the following three regions:
\begin{enumerate}
\item[1.] Low temperature: \; $aT\ll rT\ll 1$,
\item[2.] Medium temperature:\; $aT\ll 1\ll rT$,
\item[3.] High temperature:\; $1\ll aT\ll rT$.
\end{enumerate}

\subsection{The low temperature region}
In the low temperature region, the Casimir force is dominated by the zero temperature term. Substitute the zero temperature Casimir force density \eqref{eq4_7_1}
into \eqref{eq4_6_3}, we find that the proximity force approximation of the zero temperature Casimir force is
\begin{equation*}
\begin{split}
F_{\text{Cas}}^{ T=0,\text{PFA}}=&  -\frac{\pi^2Lr}{240}\int_0^1\frac{1}{(a+rt)^4}(1-t)\frac{dt}{\sqrt{2t-t^2}}.\end{split}
\end{equation*}Making a change of variables $t\mapsto au$, we find that
\begin{equation*}
\begin{split}
F_{\text{Cas}}^{ T=0,\text{PFA}}= &-\frac{\pi^2Lra}{240 }\int_0^{\frac{1}{a}}\frac{1}{(a+aru)^4 }\frac{1-au}{\sqrt{2au-a^2u^2} }du \\\sim  &-\frac{\pi^2Lr}{240 \sqrt{2}a^{\frac{7}{2}}}\int_0^{\infty}\frac{1}{(1+ru)^4\sqrt{u}}du
= -\frac{\pi^3L\sqrt{r}}{768\sqrt{2}a^{\frac{7}{2}}}.
\end{split}
\end{equation*}This agrees with the leading term found by using exact method in \cite{1}. For the thermal correction, the leading term in the low temperature region comes from the distance independent term  \eqref{eq4_7_2}, which gives
\begin{equation}\label{eq4_13_1}
\begin{split}
\Delta_TF_{\text{Cas}}^{ \text{PFA}}\sim   -\frac{\pi^2T^4Lr}{45}\int_0^1 (1-t)\frac{dt}{\sqrt{2t-t^2}}
=  -\frac{\pi^2T^4Lr}{45}.
\end{split}
\end{equation}For the remaining terms in \eqref{eq4_6_1}, since $aT\ll rT\ll 1$,
$$\exp\left(-\frac{1}{(a+rt)T}\right)\ll 1$$ for any $0\leq t\leq 1$. Therefore, they give exponentially small contributions.
In other words, we find that in the low temperature region, the proximity force approximation gives
\begin{equation}\label{eq5_31_1}F_{\text{Cas}}^{\text{PFA}}=-\frac{\pi^3L\sqrt{r}}{768\sqrt{2}a^{\frac{7}{2}}}+\ldots,\end{equation} and the leading term of the thermal correction is \begin{equation}\label{eq6_6_1}\Delta_TF_{\text{Cas}}^{\text{PFA}}\sim-\frac{\pi^2T^4Lr}{45}+\ldots.\end{equation}Notice that this term is independent of $a$. In the same way, we find that for the Casimir energy,
$$E_{\text{Cas}}^{\text{PFA}}=-\frac{\pi^3L\sqrt{r}}{1920\sqrt{2}a^{\frac{5}{2}}}+\ldots.$$For the thermal correction to the Casimir energy, we have
\begin{equation*}
\begin{split}
\Delta_TE_{\text{Cas}}^{\text{PFA}}\sim \frac{ \pi^2LrT^4}{45}\int_0^1 (a+rt)(1-t)\frac{dt}{\sqrt{2t-t^2}}= \frac{ \pi^2Lr^2T^4}{45}\left(1-\frac{\pi}{4}\right)+\frac{\pi^2T^4Lar}{45}.
\end{split}
\end{equation*}The first term is independent of $a$ and so it does not contribute to the Casimir force. The second term gives rise to the leading contribution to the thermal correction of the Casimir force \eqref{eq6_6_1}.

\subsection{The medium temperature region}
In the medium temperature region, one cannot ignore the contribution to the proximity force from the exponential terms in \eqref{eq4_6_1}. In fact, when $aT\ll 1 \ll rT$, the function
$$\exp\left(-\frac{1}{(a+rt)T}\right) $$ is small if $t\rightarrow 0$ but it becomes $\sim 1$ when $t\rightarrow 1$. The contribution of these exponential terms to the proximity force approximation is given by
\begin{equation*}\begin{split}
I =&  \pi Lr T  \sum_{k=1}^{\infty}\sum_{l=1}^{\infty} \frac{k^2}{l}\int_0^1 \frac{1}{(a+rt)^3}\exp\left(-\frac{\pi kl}{(a+rt)T}\right)(1-t)\frac{dt}{\sqrt{2t-t^2}}\\
 \xrightarrow{a\rightarrow 0}&\frac{\pi L  T  }{  r^2}\sum_{k=1}^{\infty}\sum_{l=1}^{\infty} \frac{k^2}{l}\int_0^1 \frac{1}{t^3}\exp\left(-\frac{\pi kl}{rTt}\right)(1-t)\frac{dt}{\sqrt{2t-t^2}}\\
 =&\frac{\pi L  T  }{  r^2}\sum_{k=1}^{\infty}\sum_{l=1}^{\infty} \frac{k^2}{l}\int_1^{\infty}  \exp\left(-\frac{\pi kl}{rT }t\right)\frac{t(t-1)}{\sqrt{2t-1}}dt.
\end{split}
\end{equation*}
To find the asymptotic behavior of this term when $rT\gg 1$, we use the inverse Mellin transform formula
\begin{equation}\label{eq6_1_4}e^{-v}=\frac{1}{2\pi i}\int_{c-i\infty}^{c+i\infty}   \Gamma(z) v^{-z}dz,\end{equation} which gives
\begin{equation}\label{eq6_1_1}
\begin{split}
I= &\frac{\pi L  T  }{ r^2   }\frac{1}{2\pi i}\int_{c-i\infty}^{c+i\infty} \Gamma(z)\left(\frac{rT    }{\pi}\right)^{z}
\zeta_R(z-2)\zeta_R(z+1)\mathcal{A}(z)dz,
\end{split}
\end{equation}where $\mathcal{A}(z)$ is the function
\begin{equation*}
\mathcal{A}(z)=\int_{1}^{\infty} \frac{t^{1-z}(t-1)}{\sqrt{2t-1}}dt.
\end{equation*}A change of variables $t\mapsto (t+1)/2$ gives
\begin{equation*}\begin{split}
\mathcal{A}(z)=&2^{z-3}\int_1^{\infty}(t+1)^{1-z}(t-1)\frac{dt}{\sqrt{t}}\\
=&2^{z-3}\left(\int_0^{\infty}\frac{t^{\frac{1}{2}}}{(t+1)^{z-1}}dt-\int_0^{\infty}\frac{t^{-\frac{1}{2}}}{(t+1)^{z-1}}dt+\int_0^1(t+1)^{1-z}(1-t)\frac{dt}{\sqrt{t}}\right)\\
=&2^{z-3}\left(-\sqrt{\pi}(z-3)\frac{\Gamma\left(z-\frac{5}{2}\right)}{\Gamma(z-1)}+\int_0^1(t+1)^{1-z}(1-t)\frac{dt}{\sqrt{t}}\right).
\end{split}\end{equation*}The last integral is convergent for any $z$. Therefore, $\mathcal{A}(z)$ has poles at $z= 5/2, 3/2,1/2,-1/2,\ldots$. Moreover, for $j=0,1,2,\ldots$,
\begin{align*}
\text{Res}_{z=\frac{5}{2}-j}\left(\Gamma(z)\mathcal{A}(z)\right)= \frac{\sqrt{\pi}}{2^{j+\frac{1}{2}}}\frac{(-1)^j\left(j+\frac{1}{2}\right)\left(\frac{3}{2}-j\right)}{j! }.
\end{align*}
The integrand in \eqref{eq6_1_1} has simple poles at $z= 5/2, 3/2,1/2,-1/2,\ldots$ and at $z=0$ and $z=3$. Evaluate the residues gives the asymptotic behavior for $I$ when $rT\gg1$. Namely:
\begin{equation*}
\begin{split}
I\sim & \frac{\pi^2 L     }{ r^3   }\left(2\left(\frac{rT    }{\pi}\right)^{4}\zeta_R(4)\mathcal{A}(3)+\zeta_R(-2)\left(\frac{rT    }{\pi}\right) \lim_{z\rightarrow 0}\Gamma(z)\mathcal{A}(z)
\right.\\&\hspace{1cm}\left.+\sum_{j=0}^{\infty} \left(\frac{rT    }{\pi}\right)^{\frac{7}{2}-j}
\zeta_R\left(\frac{1}{2}-j\right)\zeta_R\left(\frac{7}{2}+j\right)\frac{\sqrt{\pi}}{2^{j+\frac{1}{2}}}\frac{(-1)^j\left(j+\frac{1}{2}\right)\left(\frac{3}{2}-j\right)}{j! }\right).
\end{split}
\end{equation*} Since $\mathcal{A}(3)=1$ and $\zeta_R(-2)=0$, we find that
\begin{equation*}
\begin{split}
I\sim & \frac{\pi^2 Lr T^4}{45}+  LT^{\frac{7}{2}} \sqrt{r} \frac{3 }{4\sqrt{2}\pi}\zeta_R\left(\frac{1}{2}\right)\zeta_R\left(\frac{7}{2}\right)- \frac{LT^{\frac{5}{2}} }{\sqrt{r}} \frac{3 }{8\sqrt{2} }\zeta_R\left(-\frac{1}{2}\right)\zeta_R\left(\frac{9}{2}\right)- \frac{\pi LT^{\frac{3}{2}} }{r^{\frac{3}{2}}} \frac{5 }{32\sqrt{2} }\zeta_R\left(-\frac{3}{2}\right)\zeta_R\left(\frac{11}{2}\right)+\ldots
\end{split}\end{equation*}The first term, $\displaystyle \frac{\pi^2 Lr T^4}{45}$, cancel with the term \eqref{eq4_13_1} coming from \eqref{eq4_7_2}. Therefore we find that if $aT\ll 1\ll rT$, the asymptotic behavior of the  thermal correction is
$$ \Delta_TF_{\text{Cas}}^{\text{PFA}}\sim LT^{\frac{7}{2}} \sqrt{r} \frac{3 }{4\sqrt{2}\pi}\zeta_R\left(\frac{1}{2}\right)
\zeta_R\left(\frac{7}{2}\right)- \frac{LT^{\frac{5}{2}} }{\sqrt{r}} \frac{3 }{8\sqrt{2} }\zeta_R\left(-\frac{1}{2}\right)\zeta_R\left(\frac{9}{2}\right)- \frac{\pi LT^{\frac{3}{2}} }{r^{\frac{3}{2}}} \frac{5 }{32\sqrt{2} }\zeta_R\left(-\frac{3}{2}\right)\zeta_R\left(\frac{11}{2}\right)+\ldots.  $$In particular, the leading term is
\begin{equation}\label{eq6_2_16}\Delta_TF_{\text{Cas}}^{\text{PFA}}\sim LT^{\frac{7}{2}} \sqrt{r} \frac{3 }{4\sqrt{2}\pi}\zeta_R\left(\frac{1}{2}\right)
\zeta_R\left(\frac{7}{2}\right)\sim -0.2778LT^{\frac{7}{2}} \sqrt{r}.\end{equation}
This agree with the result obtained in \cite{6}. The leading term of the Casimir force is still given by the zero temperature term \eqref{eq5_31_1} since $aT\ll 1$.

For the Casimir energy, a similar computation computation gives  
\begin{equation}\label{eq6_2_23}\Delta_TE_{\text{Cas}}^{\text{PFA}}\sim -\frac{L\sqrt{r}T^{\frac{5}{2}}}{4\sqrt{2}\pi}\zeta_R\left(\frac{5}{2}\right)\zeta_R\left(\frac{3}{2}\right)
+\frac{3L r^{-\frac{1}{2}}T^{\frac{3}{2}}}{32\sqrt{2}\pi}\zeta_R\left(\frac{3}{2}\right)\zeta_R\left(\frac{5}{2}\right)+\ldots, \end{equation}  which is independent of $a$. In particular, the leading term of the thermal correction to the Casimir energy is
 \begin{equation}\label{eq6_2_15}\Delta_TE_{\text{Cas}}^{\text{PFA}}\sim -\frac{L\sqrt{r}T^{\frac{5}{2}}}{4\sqrt{2}\pi}\zeta_R\left(\frac{5}{2}\right)\zeta_R\left(\frac{3}{2}\right)=-0.1972LT^{\frac{5}{2}} \sqrt{r}. \end{equation}
\subsection{High temperature region}

In the high temperature region, the Casimir force is dominated by the classical term. Substitute the term \eqref{eq4_7_3} into \eqref{eq4_6_3},   the proximity force approximation gives
\begin{equation*}
F_{\text{Cas}}^{\text{PFA}}= -\frac{\zeta_R(3)Lr T  }{4\pi  }\int_0^1 \frac{1}{(a+rt)^3} (1-t)\frac{dt}{\sqrt{2t-t^2}}.\end{equation*}
Making a change of variable $t\mapsto au$, we obtain
\begin{equation*}
\begin{split}
F_{\text{Cas}}^{\text{PFA}}= &-\frac{\zeta_R(3)Lr T  a}{4\pi  }\int_0^{\frac{1}{a}} \frac{1}{(a+rau)^3} (1-au)\frac{du}{\sqrt{2au-a^2u^2}}\\ \sim & -\frac{\zeta_R(3)Lr T  }{4\sqrt{2}\pi a^{\frac{5}{2}}  }\int_0^{\infty}\frac{1}{(1+ru)^3\sqrt{u}}du
=-\frac{3\zeta_R(3)L\sqrt{r} T  }{32\sqrt{2}  a^{\frac{5}{2}}  }.
\end{split}
\end{equation*}
For the remaining terms in \eqref{eq4_6_2}, since
$$\exp\left(-4\pi(a+rt)T\right)$$ is exponentially small when $1\ll aT\ll rT$, they give exponentially small contribution to the proximity force approximation. In other words, in the high temperature region, the proximity force approximation gives
\begin{equation*}
F_{\text{Cas}}^{\text{PFA}}=-\frac{3\zeta_R(3)L\sqrt{r} T  }{32\sqrt{2}  a^{\frac{5}{2}}  }+\ldots.
\end{equation*}For the Casimir energy, one has
\begin{equation}\label{eq6_2_17}
E_{\text{Cas}}^{\text{PFA}}=-\frac{ \zeta_R(3)L\sqrt{r} T  }{16\sqrt{2}  a^{\frac{3}{2}}  }+\ldots.
\end{equation}

The results obtained above are for a cylinder above a plate with Dirichlet or Neumann boundary conditions on both the cylinder and the plate. For perfectly conducting cylinder and   plate, all the results have to be multiplied by two due to the TE and TM modes.

\section{Exact finite temperature Casimir energy}
In this section, we consider exact expressions for the Casimir energy of the cylinder-plate system. The exact zero temperature Casimir interaction energy of the cylinder-plate system is given by
\begin{equation}\label{eq4_15_1}
\begin{split}
E_{\text{Cas}}^{T=0}=\frac{L}{2\pi^2}\int_0^{\infty}\int_0^{\infty} \text{Tr}\ln \left(\mathbf{1}-\mathbf{M}\left(\sqrt{\xi^2+k^2}\right)\right)d\xi dk=\frac{L}{4\pi}\int_0^{\infty}\xi \text{Tr}\ln \left(\mathbf{1}-\mathbf{M}(\xi)\right)d\xi,
\end{split}
\end{equation}where $\mathbf{M}(\xi)$ is an infinite matrix with elements $M_{jk}(\xi),\;-\infty<j,k<\infty$,
\begin{equation*}
M_{jk}(\xi)=M_{jk}^D(\xi)=\frac{I_k(r\xi)}{K_j(r\xi)}K_{j+k}(2H\xi)
\end{equation*}for Dirichlet boundary conditions, and
\begin{equation*}
M_{jk}(\xi)=M_{jk}^N(\xi)=-\frac{I_k'(r\xi)}{K_j'(r\xi)}K_{j+k}(2H\xi)
\end{equation*}
for Neumann boundary conditions.
For perfectly conducting boundary conditions, the Casimir energy is the sum of the Casimir energy for Dirichlet boundary conditions and the Casimir energy for Neumann boundary conditions. These formulas have been proved by using the mode summation approach \cite{2} and the path integral approach \cite{1,3}.

Using the Matsubara formalism, one then finds that the finite temperature Casimir energy is given by
\begin{equation}\label{eq4_18_1}
\begin{split}
E_{\text{Cas}}=& \frac{TL}{\pi} \sum_{l=0}^{\infty}\!'\int_0^{\infty} \text{Tr}\,\ln \left(\mathbf{1}-\mathbf{M}\left(\sqrt{\xi_l^2+k^2}\right)\right) dk\\
 =&\frac{TL}{\pi} \sum_{l=0}^{\infty}\!'\int_{\xi_l}^{\infty}\frac{\xi}{\sqrt{\xi^2-\xi_l^2}}  \text{Tr}\,\ln (\mathbf{1}-\mathbf{M}(\xi))d\xi,
\end{split}
\end{equation}where $\xi_l=2\pi l T$ are the Matsubara frequencies. This formula is suitable for studying the high temperature behavior of the Casimir energy. In particular, in the high temperature region, the Casimir energy is dominated by the classical term corresponding to zero Matsubara frequency:
\begin{equation*}
\begin{split}
E_{\text{Cas}} \sim &E_{\text{Cas}}^{\text{cl}}=\frac{TL}{2\pi} \int_{0}^{\infty}  \text{Tr}\,\ln (\mathbf{1}-\mathbf{M}(\xi))d\xi.
\end{split}
\end{equation*}

To study the low temperature behavior, one can use the Poisson summation formula
\begin{equation*}
\sum_{l=0}^{\infty}\!'f(l)=\int_0^{\infty}f(x)dx+2\sum_{l=1}^{\infty}\int_0^{\infty}f(x)\cos 2\pi l x dx
\end{equation*} with
\begin{equation*}
\begin{split}
f(x)=\frac{TL}{\pi}  \int_{2\pi x T}^{\infty}\frac{\xi}{\sqrt{\xi^2-(2\pi xT)^2}}  \text{Tr}\,\ln (\mathbf{1}-\mathbf{M}(\xi))d\xi.
\end{split}
\end{equation*}Then the term $\displaystyle \int_0^{\infty}f(x)dx$ reproduces the zero temperature Casimir energy \eqref{eq4_15_1}. The thermal correction to the Casimir energy is
\begin{equation}\label{eq5_31_2}
\begin{split}
\Delta_T E_{\text{Cas}} =2\int_0^{\infty}f(x)\cos 2\pi l x dx=&\frac{2TL}{2\pi^2 T}\int_0^{\infty}\int_x^{\infty}\frac{\xi}{\sqrt{\xi^2-x^2}}  \text{Tr}\,\ln (\mathbf{1}-\mathbf{M}(\xi))d\xi \cos\frac{lx}{T} dx\\
=&\frac{L}{\pi^2}\int_0^{\infty} \xi\int_0^{\xi}\frac{1}{\sqrt{\xi^2-x^2}} \cos\frac{lx}{T} dx \text{Tr}\,\ln (\mathbf{1}-\mathbf{M}(\xi))d\xi \\
=&\frac{L}{2\pi}\int_0^{\infty} \xi J_0\left(\frac{l\xi}{T} \right)\text{Tr}\,\ln (\mathbf{1}-\mathbf{M}(\xi))d\xi.
\end{split}\end{equation}Therefore, the low temperature expansion for the Casimir energy is given by
\begin{equation*}
E_{\text{Cas}}=E_{\text{Cas}}^{T=0}+\frac{L}{2\pi}\sum_{l=1}^{\infty}\int_0^{\infty} \xi J_0\left(\frac{l\xi}{T} \right)\text{Tr}\,\ln (\mathbf{1}-\mathbf{M}(\xi))d\xi.
\end{equation*}Alternatively, one can apply  the Abel-Plana summation formula to the first line of \eqref{eq4_18_1} to obtain\begin{equation*}
\begin{split}
E_{\text{Cas}}=& E_{\text{Cas}}^{T=0}+\frac{L}{2\pi^2}\int_0^{\infty}\int_k^{\infty}\frac{i\left( \text{Tr}\, \ln (\mathbf{1}-\mathbf{M}(i\sqrt{\xi^2-k^2}))- \text{Tr}\, \ln (\mathbf{1}-\mathbf{M}(-i\sqrt{\xi^2-k^2}))\right)}{e^{\frac{\xi}{T}}-1}d\xi dk.
\end{split}
\end{equation*}The second term is the thermal correction $\Delta_TE_{\text{Cas}}$. By a change of variables, it is equal to
\begin{equation*}
\begin{split}
\Delta_TE_{\text{Cas}}=&\frac{L}{2\pi^2}\int_0^{\infty}\int_0^{\infty}\frac{i\left( \text{Tr}\, \ln (\mathbf{1}-\mathbf{M}(i\xi))- \text{Tr}\, \ln (\mathbf{1}-\mathbf{M}(-i\xi))\right)}{e^{\frac{\sqrt{\xi^2+k^2}}{T}}-1}\frac{\xi d\xi}{\sqrt{\xi^2+k^2}} dk.
\end{split}
\end{equation*}Since
\begin{equation*}
\begin{split}
\int_0^{\infty}\frac{1}{e^{\frac{\sqrt{\xi^2+k^2}}{T}}-1}\frac{dk}{\sqrt{\xi^2+k^2}} =&\int_{\xi}^{\infty}\frac{1}{e^{\frac{u}{T}}-1}\frac{du}{\sqrt{u^2-\xi^2}}
= \sum_{l=1}^{\infty}\int_{\xi}^{\infty}e^{-\frac{lu}{T}}\frac{du}{\sqrt{u^2-\xi^2}}
= \sum_{l=1}^{\infty} K_0\left(\frac{l\xi}{T}\right),
\end{split}
\end{equation*}we find that
\begin{equation*}
\begin{split}
\Delta_TE_{\text{Cas}}=& \frac{L}{2\pi^2}\sum_{l=1}^{\infty}\int_0^{\infty} \Bigl\{i\bigl[ \text{Tr}\, \ln (\mathbf{1}-\mathbf{M}(i\xi))- \text{Tr}\, \ln (\mathbf{1}-\mathbf{M}(-i\xi))\bigr]\Bigr\}\xi K_0\left(\frac{l\xi}{T}\right) d\xi.
\end{split}
\end{equation*}Notice that this term can be obtained from \eqref{eq5_31_2} by a rotation $\xi\rightarrow i\xi$ using $J_0(z)=\text{Re}\,H_0^{(1)}(z)$ and $\displaystyle H_0^{(1)}(iz)=-\frac{2i}{\pi}K_0(z)$. The integrand does not have singularities since we are considering an open geometry whose spectrum is purely continuous \cite{5}.

For the Casimir force, one has
\begin{equation*}
\begin{split}
F_{\text{Cas}}=-\frac{\pa E_{\text{Cas}}}{\pa a}=& - \frac{2TL}{\pi} \sum_{l=0}^{\infty}\!'\int_{\xi_l}^{\infty}\frac{\xi^2}{\sqrt{\xi^2-\xi_l^2}}  \text{Tr}\,\left\{ (\mathbf{1}-\mathbf{M}(\xi))^{-1}\mathbf{Q}(\xi)\right\}d\xi,
\end{split}
\end{equation*}where
\begin{equation*}
Q_{jk}(\xi)=Q_{jk}^D(\xi)=-\frac{I_k(r\xi)}{K_j(r\xi)}K_{j+k}'(2H\xi)
\end{equation*}for Dirichlet boundary conditions, and
\begin{equation*}
Q_{jk}(\xi)=Q_{jk}^N(\xi)=\frac{I_k'(r\xi)}{K_j'(r\xi)}K_{j+k}'(2H\xi)
\end{equation*}for Neumann boundary conditions. Since
$M_{jk}(\xi)$ and $Q_{jk}(\xi)$ are positive functions, using $(\mathbf{1}-\mathbf{M})^{-1}=\mathbf{1}+\mathbf{M}+\mathbf{M}^2+\ldots$, one concludes that
the Casimir force is always attractive at any temperature.
\section{Asymptotic behavior of the Casimir energy and Casimir force at small separation}
Define the dimensionless parameter $$\vep=\frac{a}{r}.$$ The first two leading terms of the zero temperature Casimir energy at small separation $\vep\ll 1$ has been obtained by Bordag in \cite{1}. It was found that for Dirichlet conditions on both the cylinder and the plate,
\begin{equation}\label{eq6_2_20}
E^{D,T=0}_{\text{Cas}}=-\frac{\pi^3L\sqrt{r}}{1920\sqrt{2}a^{\frac{5}{2}}}\left(1+\frac{7}{36}\vep+\ldots\right);
\end{equation}whereas for Neumann conditions on both the cylinder and the plate,
\begin{equation}\label{eq6_2_21}
E^{N,T=0}_{\text{Cas}}=-\frac{\pi^3L\sqrt{r}}{1920\sqrt{2}a^{\frac{5}{2}}}\left(1+\left[\frac{7}{36}-\frac{40}{3\pi^2}\right]\vep+\ldots\right).
\end{equation}For the Casimir force, one then obtains
\begin{equation*}\begin{split}
F^{D,T=0}_{\text{Cas}}=&-\frac{\pi^3L\sqrt{r}}{768\sqrt{2}a^{\frac{5}{2}}}\left(1+\frac{7}{60}\vep+\ldots\right),\\
F^{N,T=0}_{\text{Cas}}=&-\frac{\pi^3L\sqrt{r}}{768\sqrt{2}a^{\frac{5}{2}}}\left(1+\left[\frac{7}{60}-\frac{8}{\pi^2}\right]\vep+\ldots\right).\end{split}\end{equation*}

In the following, we would  derive the asymptotic behavior of the Casimir energy and the Casimir force at finite temperature. This has been studied numerically in \cite{6} using worldline method. The approach   used here is similar to the one used in \cite{1} for the zero temperature case. The leading asymptotic behavior would be computed analytically using the exact formula for the Casimir energy and compared to the proximity force approximation. We would also discuss the limitation of this method when trying to obtain the first order correction analytically.

The starting point is the formula \eqref{eq4_18_1}. Making a change of variable $\xi=\omega/r$, one obtains
\begin{equation*}
\begin{split}
E_{\text{Cas}}=&  \frac{TL}{\pi r} \sum_{l=0}^{\infty}\!'\int_{r\xi_l}^{\infty}\frac{\omega}{\sqrt{\omega^2-r^2\xi_l^2}}  \text{Tr}\,\ln (\mathbf{1}-\mathbf{A}(\omega))d\omega,
\end{split}
\end{equation*}where
\begin{equation*}
A_{jk}^D(\omega)=\frac{I_k(\omega)}{K_j(\omega)}K_{j+k}(2\omega(1+\vep))
,\hspace{1cm}
A_{jk}^N(\omega)=-\frac{I_k'(r\omega)}{K_j'(r\omega)}K_{j+k}(2\omega(1+\vep)).
\end{equation*}
Expanding the logarithm, we find that
\begin{equation*}
\begin{split}
E_{\text{Cas}}=&  -\frac{TL}{\pi r} \sum_{s=0}^{\infty}\frac{1}{s+1}\sum_{l=0}^{\infty}\!'\int_{r\xi_l}^{\infty}\frac{\omega}{\sqrt{\omega^2-r^2\xi_l^2}}  \sum_{j_0=-\infty}^{\infty}\ldots\sum_{j_s=-\infty}^{\infty}\prod_{i=0}^sA_{j_i,j_{i+1}}(\omega)d\omega.
\end{split}
\end{equation*}
Debye asymptotic expansions
of Bessel functions give an expansion of $A_{\nu_1\nu_2}(\omega)$ in the form:
\begin{equation}\label{eq5_31_4}
\begin{split}
A_{\nu_1\nu_2}^D(\omega)\sim &
\frac{1}{\sqrt{2\pi}}\sqrt{\frac{\nu_1}{\nu_2\nu_3}}
\left(\frac{1+\omega_1^2}{(1+\omega_2)^2(1+\omega_3)^2}\right)^{\frac{1}{4}}\exp\Bigl(-\nu_3\eta(\omega_3)+\nu_1\eta(\omega_1)+\nu_2\eta(\omega_2)\Bigr)
\\&\times\left(1+\frac{u_1(t(\omega_1))}{\nu_1}+\frac{u_1(t(\omega_2))}{\nu_2}-\frac{u_1(t(\omega_3))}{\nu_3}\right),
\end{split}
\end{equation}\begin{equation}\label{eq6_6_4}
\begin{split}A_{\nu_1 \nu_2}^N(\omega)\sim &
\frac{1}{\sqrt{2\pi}}\sqrt{\frac{\nu_2}{\nu_1\nu_3}}
\left(\frac{1+\omega_2^2}{(1+\omega_1)^2(1+\omega_3)^2}\right)^{\frac{1}{4}}\exp\Bigl(-\nu_3\eta(\omega_3)+\nu_1\eta(\omega_1)+\nu_2\eta(\omega_2)\Bigr)
\\&\times\left(1+\frac{v_1(t(\omega_1))}{\nu_1}+\frac{v_1(t(\omega_2))}{\nu_2}-\frac{u_1(t(\omega_3))}{\nu_3}\right),
\end{split}
\end{equation}where
\begin{equation*}
\begin{split}
\omega_1=&\frac{\omega}{\nu_1},\quad \omega_2=\frac{\omega}{\nu_2},\quad\omega_3=\frac{2 (1+\vep)\omega}{\nu_3}
\\
\eta(z)=&\sqrt{1+z^2}+\log\frac{z}{1+\sqrt{1+z^2}},\hspace{1cm}t(z)=\frac{1}{\sqrt{1+z^2}},\\
u_1(t)=&-\frac{5t^3-3t}{24},\hspace{1cm}v_1(t)=\frac{7t^3-9t}{24}.
\end{split}\end{equation*}
Observe that after interchanging the subscripts $1$ and $2$ in the first line of \eqref{eq6_6_4}, it coincides with the first line of \eqref{eq5_31_4}.

One can show  that the expression $-\nu_3\eta(\omega_3)+\nu_1\eta(\omega_1)+\nu_2\eta(\omega_2)$  has a maximum value of 0 when $\nu_1=\nu_2$ and $\vep=0$. This suggest that we rename $j_0$ as $m$, and $j_i$ as $m+n_i$ for $1\leq i\leq s$, and treat $n_1,\ldots,n_s$ as perturbed variables. Making a change of variables $\omega\mapsto m\omega$, and replacing the summation over $n_i$ to integration (which is the leading term  after applying Poisson resummation), we find that
\begin{equation}\label{eq5_31_6}
\begin{split}
E_{\text{Cas}}\sim &  -\frac{2TL}{\pi r} \sum_{s=0}^{\infty}\frac{1}{s+1}\sum_{l=0}^{\infty}\!'\sum_{m=0}^{\infty}\!'m\int_{\frac{r\xi_l}{m}}^{\infty}\frac{\omega}{\sqrt{\omega^2-\frac{r^2\xi_l^2}{m^2}}}  \int_{ -\infty}^{\infty}\ldots\int_ {-\infty}^{\infty}\prod_{i=0}^sA_{m+n_i,m+n_{i+1}}(m\omega)dn_1\ldots dn_sd\omega.
\end{split}
\end{equation}Treating $\vep$ and $n_1,\ldots, n_s$ as perturbed variables, one can expand $A_{m+n_i,m+n_{i+1}}(m\omega)$ with the help of computer.
In fact, for the    term $-\nu_3\eta(\omega_3)+\nu_1\eta(\omega_1)+\nu_2\eta(\omega_2)$, where $\nu_1=m+n_i,\nu_2=m+n_{i+1},\nu_3=2m+n_i+n_{i+1}$, we have an expansion of the form
$$-\nu_3\eta(\omega_3)+\nu_1\eta(\omega_1)+\nu_2\eta(\omega_2)\sim \sum_{k=0}^{N_1}\sum_{j=0}^{N_2}\vep^j m^{1-k}\mathscr{F}_{kj}(n_i,n_{i+1})\mathscr{G}_{kj}(\omega),$$
where $\mathscr{F}_{kj}(n_i,n_{i+1})$ is a homogeneous polynomial of degree $k$ in $n_i$ and $n_{i+1}$.
The leading terms are
$$-\nu_3\eta(\omega_3)+\nu_1\eta(\omega_1)+\nu_2\eta(\omega_2)\sim -2\vep m \sqrt{1+\omega^2}-  \frac{ (n_i-n_{i+1})^2}{4m\sqrt{1+\omega^2}}.$$
From here, we see that the leading contribution comes from $m\sim  \vep^{-1}$, $n_i, n_{i+1}\sim m^{1/2}\sim \vep^{-1/2}$. Consequently, the term $\vep^jm^{1-k}\mathscr{F}_{kj}(n_i,n_{i+1})$ is of order $\vep^{\frac{k}{2}+j-1}$. To keep everything to the order $\vep$, we need to expand up to the terms with $j+k/2\leq 2$. Doing similar order analysis to the other terms, and expanding the terms that are of order $\sqrt{\vep}$ and $\vep$ in the exponential using $e^x=1+x+x^2/2+\ldots$, we find that
\begin{equation}\label{eq5_31_5}\begin{split}
A_{m+ n_i,m+ n_{i+1}}(m\omega)\sim & \frac{1}{2\sqrt{\pi m}(1+\omega^2)^{\frac{1}{4}}}\exp\left(-2\vep m \sqrt{1+\omega^2}-  \frac{ (n_i-n_{i+1})^2}{4m\sqrt{1+\omega^2}}\right)
\left(1+ c(n_i,n_{i+1})+ b(n_i,n_{i+1})+\ldots\right),\end{split}
\end{equation}where $ c(n_i,n_{i+1})$ and $ b(n_i,n_{i+1})$ are terms of order $\sqrt{\vep}$ and $\vep$ respectively. In the case of Dirichlet boundary conditions,
\begin{align*}
c^D(n_{i},n_{i+1}) =& \frac{ n_i-3n_{i+1} }{4m(1+\omega^2)}+   \frac{(n_i+n_{i+1})(n_i-n_{i+1})^2}{8m^2(1+\omega^2)^{\frac{3}{2}}}
-\vep \frac{1}{\sqrt{1+\omega^2}}(n_i+n_{i+1})\\
b^D(n_{i},n_{i+1})=&-\frac{ (5n_i^2 + 2n_in_{i+1} - 19n_{i+1}^2) -\omega^2( 6n_i^2 - 4n_in_j - 10n_{i+1}^2)}{32m^2(1 + \omega^2)^2 }-\vep\frac{ \omega^2}{2 (1+\omega^2)}\\
&+\frac{ (n_i-3n_{i+1})(n_i+n_{i+1})(n_i-n_{i+1})^2}{32m^3(1+\omega^2)^{\frac{5}{2}}}-\vep\frac{ (n_i-3n_{i+1})(n_i+n_{i+1})}{4m(1+\omega^2)^{\frac{3}{2}}}\\
&+   \frac{(n_i+n_{i+1})^2(n_i-n_{i+1})^4}{128m^4(1+\omega^2)^{3}}+\vep^2 \frac{1}{2(1+\omega^2)}(n_i+n_{i+1})^2-\vep\frac{1}{8m^2(1+\omega^2)^2}(n_i+n_{i+1})^2(n_i-n_{i+1})^2\\
&-\vep \omega^2\frac{(n_i+n_{i+1})^2}{4m(1+\omega^2)^{\frac{3}{2}}}-  (2 -\omega^2)\frac{(n_i-n_{i+1})^2(7n_i^2+10n_in_{i+1}+7n_{i+1}^2)}{192m^3(1+\omega^2)^{\frac{5}{2}}}
+\vep^2 \frac{m }{\sqrt{1+\omega^2}}-\frac{2  - 3\omega^2 }{16m(1+\omega^2)^{\frac{3}{2}}}
\end{align*}Only the last term of $b^D(n_{i},n_{i+1})$ comes from the second line of \eqref{eq5_31_4}. Therefore, for the case of Neumann boundary conditions, a comparison between \eqref{eq5_31_4} and \eqref{eq6_6_4} shows that\begin{equation}\label{eq6_1_2}
c^N(n_i,n_{i+1})=c^D(n_{i+1},n_i),\hspace{1cm} b^N(n_i,n_{i+1})=b^D(n_{i+1},n_i)-\frac{\omega^2}{m(1+\omega^2)^{\frac{3}{2}}}.
\end{equation}These relations between $c^N$ and $c^D$, and between $b^N$ and $b^D$   can help us to derive the asymptotic behavior for the Neumann case from the asymptotic behavior for the Dirichlet case.

Substituting \eqref{eq5_31_5} into \eqref{eq5_31_6}, we find that
\begin{equation}\label{eq5_31_7}
\begin{split}
E_{\text{Cas}}\sim &  -\frac{ TL}{\pi^{\frac{3}{2}} r} \sum_{s=0}^{\infty}\frac{1}{(s+1)^{\frac{3}{2}}} \sum_{l=0}^{\infty}\!'\sum_{m=0}^{\infty}\!'m^{ \frac{ 1}{2}}
\int_{\frac{r\xi_l}{m}}^{\infty}\frac{\omega}{\sqrt{\omega^2-\frac{r^2\xi_l^2}{m^2}}}  (1+\omega^2)^{-\frac{ 1}{4}}\exp\left(-2\vep (s+1) m \sqrt{1+\omega^2}\right)\left(1+\mathscr{A}^s\right)d\omega,
\end{split}
\end{equation}
where
\begin{equation*}\begin{split}
\mathscr{A}^s=&\sum_{i=0}^s\mathscr{B}^s_i+\sum_{i=0}^{s-1}\sum_{j=i+1}^s\mathscr{C}_{ij}^s,\\
\mathscr{B}^s_i=&  \frac{\sqrt{s+1}}{2^s\pi^{\frac{s}{2}}m^{\frac{s}{2}}(1+\omega^2)^{\frac{s}{4}}}\int_{-\infty}^{\infty}\ldots\int_{-\infty}^{\infty}\exp\left( -  \frac{ (n_i-n_{i+1})^2}{4m\sqrt{1+\omega^2}}\right)b(n_i,n_{i+1})dn_1\ldots dn_s,\\
\mathscr{C}^s_{ij}=&  \frac{\sqrt{s+1}}{2^s\pi^{\frac{s}{2}}m^{\frac{s}{2}}(1+\omega^2)^{\frac{s}{4}}}\int_{-\infty}^{\infty}\ldots\int_{-\infty}^{\infty}\exp\left( -  \frac{ (n_i-n_{i+1})^2}{4m\sqrt{1+\omega^2}}\right)c(n_i,n_{i+1})c(n_j,n_{j+1})dn_1\ldots dn_s.
\end{split}\end{equation*}We have used the fact that
$$\int_{-\infty}^{\infty}\ldots\int_{-\infty}^{\infty}\exp\left( -  \frac{ (n_i-n_{i+1})^2}{4m\sqrt{1+\omega^2}}\right) dn_1\ldots dn_s=
\frac{2^s\pi^{\frac{s}{2}}m^{\frac{s}{2}}(1+\omega^2)^{\frac{s}{4}}}{\sqrt{s+1}}.$$
$\mathscr{A}^s$ is the term of order $\vep$. The term with order $\sqrt{\vep}$ has dropped out since $c(n_i,n_{i+1})$ is odd in $n_i$ or $n_{i+1}$.

$\mathscr{B}^s_i$ and $\mathscr{C}^s_{ij}$ are Gaussian integrals. They can be computed in the same way as explained in \cite{1}. With the help of a machine, we find that for the case of Dirichlet boundary conditions,
\begin{equation*}\begin{split}
\mathscr{B}^{s,D}_i=& \frac{\vep}{2}\left(-(4i+2)+\frac{(2i+1)^2}{s+1}\right)+\frac{1}{16m}\left(-\frac{8i+4}{s+1}+\frac{24i^2+24i+7}{(s+1)^2}\right) \frac{1}{(1+\omega^2)^{\frac{1}{2}}}
+ \vep^2m\left(4i+2-\frac{(2i+1)^2}{s+1}\right)\frac{1}{(1+\omega^2)^{\frac{1}{2}}}\\&
+\frac{\vep }{2}\left(4i+2+\frac{-4i^2+4i+2}{s+1}-\frac{3(2i+1)^2}{(s+1)^2}\right)\frac{1}{1+\omega^2}
+\frac{1}{16m}\left(\frac{8i+4}{s+1}+\frac{-24i^2+12i+6}{(s+1)^2}-\frac{15(2i+1)^2}{(s+1)^3}\right)\frac{1}{(1+\omega^2)^{\frac{3}{2}}},
\\
\mathscr{C}_{ij}^{s,D}=&   \vep^2m\left(8i+4-\frac{8ij+4i+4j+2}{s+1}\right) \frac{1}{(1+\omega^2)^{\frac{1}{2}}}
+ \vep  \left(  \frac{8i+4}{s+1}-\frac{12ij+6i+6j+3}{(s+1)^2}\right)\frac{1}{1+\omega^2}
\\&+\frac{1}{8m}\left( \frac{36i+18}{(s+1)^2}-\frac{60ij+30i+30j+15}{(s+1)^3}  \right)\frac{1}{(1+\omega^2)^{\frac{3}{2}}}.
\end{split}
\end{equation*}Therefore,
\begin{equation*}
\begin{split}
\mathscr{A}^{s,D}
=&  \frac{\vep^2m}{3}\Bigl( (s+1)^3 +2(s+1)\Bigr) \frac{1}{(1+\omega^2)^{\frac{1}{2}}}
+ \frac{\vep}{6}  \Bigl( (s+1)^2+2 \Bigr)\frac{1}{1+\omega^2}+\frac{1}{16m}\left(-7(s+1) +\frac{2}{s+1}    \right)\frac{1}{(1+\omega^2)^{\frac{3}{2}}} \\&-\frac{\vep}{6}\left(2(s+1)^2+1\right)
 +\frac{1}{16m}\left(4(s+1) -\frac{1}{s+1}\right)\frac{1}{(1+\omega^2)^{\frac{1}{2}}}.
\end{split}
\end{equation*}For the case of Neumann boundary conditions, the relations \eqref{eq6_1_2} suggest that we can make a change of variables $n_i\mapsto n_{s+1-i}$. Then it is easy to see that
$\mathscr{C}_{ij}^{s,N}=\mathscr{C}_{s-j,s-i}^{s,D}$ and
\begin{equation*}
\begin{split}
\mathscr{B}^{s,N}_i-\mathscr{B}^{s,D}_{s-i}= - \frac{\sqrt{s+1}}{2^s\pi^{\frac{s}{2}}m^{\frac{s}{2}}(1+\omega^2)^{\frac{s}{4}}}\int_{-\infty}^{\infty}\ldots\int_{-\infty}^{\infty}\exp\left( -  \frac{ (n_i-n_{i+1})^2}{4m\sqrt{1+\omega^2}}\right)\frac{\omega^2}{m(1+\omega^2)^{\frac{3}{2}}}dn_1\ldots dn_s=-\frac{\omega^2}{m(1+\omega^2)^{\frac{3}{2}}}.\end{split}
\end{equation*}Therefore,\begin{equation*}
\begin{split}\mathscr{A}^{s,N}-\mathscr{A}^{s,D}=-\frac{\omega^2}{m(1+\omega^2)^{\frac{3}{2}}}(s+1).
\end{split}
\end{equation*}Return to the Casimir energy \eqref{eq5_31_7}, the inverse Mellin transform formula gives
\begin{equation*}
\begin{split}
 \exp\left(-2\vep m (s+1)\sqrt{1+\omega^2}\right)=&\frac{1}{2\pi i}\int_{c-i\infty}^{c+i\infty}\Gamma(z)(2\vep)^{-z}m^{-z}(s+1)^{-z}(1+\omega^2)^{-\frac{z}{2}}dz.\end{split}
\end{equation*}
Therefore,
\begin{equation*}
\begin{split}
 &\exp\left(-2\vep m \sqrt{1+\omega^2}\right)\mathcal{A}^{s,D}\\=&\frac{1}{2\pi i}\int_{c-i\infty}^{c+i\infty}\Gamma(z+2)(2\vep)^{-z-2}m^{-z-2}(s+1)^{-z-2}(1+\omega^2)^{-\frac{z+2}{2}}\left[\frac{\vep^2m}{3}\Bigl( (s+1)^3 +2(s+1)\Bigr) \frac{1}{(1+\omega^2)^{\frac{1}{2}}}\right]dz\\
 &+\frac{1}{2\pi i}\int_{c-i\infty}^{c+i\infty}\Gamma(z+1)(2\vep)^{-z-1}m^{-z-1}(s+1)^{-z-1}(1+\omega^2)^{-\frac{z+1}{2}}\left[\frac{\vep}{6}  \Bigl( (s+1)^2+2 \Bigr)\frac{1}{1+\omega^2}-\frac{\vep}{6}\left(2(s+1)^2+1\right) \right]dz\\
  &+\frac{1}{2\pi i}\int_{c-i\infty}^{c+i\infty}\Gamma(z)(2\vep)^{-z}m^{-z}(s+1)^{-z}(1+\omega^2)^{-\frac{z}{2}}\\&\hspace{5cm}\times\left[\frac{1}{16m}\left(-7(s+1) +\frac{2}{s+1}    \right)\frac{1}{(1+\omega^2)^{\frac{3}{2}}}+\frac{1}{16m}\left(4(s+1) -\frac{1}{s+1}\right)\frac{1}{(1+\omega^2)^{\frac{1}{2}}}\right]dz\\
  =&\frac{1}{2\pi i}\int_{c-i\infty}^{c+i\infty}\Gamma(z)(2\vep)^{-z}m^{-z-1}\Biggl\{ (1+\omega^2)^{-\frac{z+3}{2}}\left[\frac{(2z+7)(2z-3)}{48}\frac{1}{(s+1)^{z-1}}
  +\frac{(2z+1)(2z+3)}{24}\frac{1}{(s+1)^{z+1}}\right]\\
&\hspace{6cm}+(1+\omega^2)^{-\frac{z+1}{2}}\left[-\frac{2z-3}{12}\frac{1}{(s+1)^{z-1}}
  -\frac{4z+3}{48}\frac{1}{(s+1)^{z+1}}\right]\Biggr\}  dz.
\end{split}
\end{equation*}Using
\begin{equation*}
\begin{split}
\int_{\frac{r\xi_l}{m}}^{\infty}\frac{\omega}{\sqrt{\omega^2-\frac{r^2\xi_l^2}{m^2}}}  \frac{1}{(1+\omega^2)^{\mu}}d\omega
=&\frac{\sqrt{\pi}}{2}\frac{m^{2\mu-1}}{(m^2+r^2\xi_l^2)^{\mu-\frac{1}{2}}}\frac{\Gamma\left(\mu-\frac{1}{2}\right)}{\Gamma(\mu)},
\end{split}
\end{equation*}we finally obtain
\begin{equation*}
\begin{split}
E_{\text{Cas}}\sim E_{\text{Cas}}^{0}+E_{\text{Cas}}^{1},
\end{split}
\end{equation*}where $E_{\text{Cas}}^{0}$ is the leading order term and $E_{\text{Cas}}^{1}$ is the first order correction term. For the case of Dirichlet boundary conditions, they are given respectively by
\begin{equation}\label{eq6_2_3}
\begin{split}
E_{\text{Cas}}^{0,D}\sim &-\frac{ TL}{2\pi  r}\sum_{s=0}^{\infty}
\frac{1}{2\pi i}\int_{c-i\infty}^{c+i\infty}\Gamma(z)(2\vep)^{-z}\frac{\mathscr{D}_1(z)}{\Gamma\left( \frac{2z+1}{4}\right)} \frac{1}{(s+1)^{z+\frac{3}{2}}}  dz,\end{split}
\end{equation}and\begin{equation}\label{eq6_2_4}
\begin{split}
E_{\text{Cas}}^{1,D}\sim &-\frac{ TL}{2\pi  r} \sum_{s=0}^{\infty}\frac{1}{2\pi i}\int_{c-i\infty}^{c+i\infty}\Gamma(z)(2\vep)^{-z} \Biggl\{ \frac{\mathscr{D}_2(z)}{\Gamma\left(\frac{2z+7}{4}\right)}\left[\frac{(2z+7)(2z-3)}{48} \frac{1}{(s+1)^{z+\frac{1}{2}}}
  +\frac{(2z+1)(2z+3)}{24} \frac{1}{(s+1)^{z+\frac{5}{2}}}\right]\\
&\hspace{6cm}+ \frac{\mathscr{D}_1(z+1)}{\Gamma\left(\frac{2z+3}{4}\right)}\left[-\frac{2z-3}{12} \frac{1}{(s+1)^{z+\frac{1}{2}}}
  -\frac{4z+3}{48} \frac{1}{(s+1)^{z+\frac{5}{2}}}\right]\Biggr\}  dz,
\end{split}
\end{equation}where
\begin{equation*}
\begin{split}
\mathscr{D}_1(z)=\Gamma\left( \frac{2z-1}{4}\right)\underbrace{\sum_{l=0}^{\infty}\!'\sum_{m=0}^{\infty}\!'}_{(l,m)\neq (0,0)} \frac{1}{(m^2+r^2\xi_l^2)^{ \frac{2z-1}{4}}},\\
\mathscr{D}_2(z)=\Gamma\left( \frac{2z+5}{4}\right) \sum_{l=0}^{\infty}\!'\sum_{m=1}^{\infty} \frac{m^2}{(m^2+r^2\xi_l^2)^{ \frac{2z+5}{4}}}.
\end{split}
\end{equation*}For the case of Neumann boundary conditions, the leading order term is the same as the Dirichlet case, i.e., $E_{\text{Cas}}^{0,N}=E_{\text{Cas}}^{0,D}$, whereas for the first order correction term,
\begin{equation}\label{eq6_2_18}
\begin{split}
E_{\text{Cas}}^{1,N}-E_{\text{Cas}}^{1,D}\sim & -\frac{ TL}{2\pi  r} \sum_{s=0}^{\infty}\frac{1}{2\pi i}\int_{c-i\infty}^{c+i\infty}\Gamma(z)(2\vep)^{-z} \Biggl\{ \frac{\mathscr{D}_2(z)}{\Gamma\left(\frac{2z+7}{4}\right)}-\frac{\mathscr{D}_1(z+1)}{\Gamma\left(\frac{2z+3}{4}\right)}\Biggr\} \frac{1}{(s+1)^{z+\frac{1}{2}}} dz.
\end{split}
\end{equation}The functions $\mathscr{D}_1(z)$ and $\mathscr{D}_2(z)$ have expansions (Chowla-Selberg formulas) of the form
\begin{equation}\label{eq6_2_1}
\begin{split}
\mathscr{D}_1(z)=& \underbrace{\frac{1}{4\sqrt{\pi}rT}\Gamma\left(\frac{ 2z-3}{4}\right)\zeta_R\left(z-\frac{3}{2}\right)}_{\mathscr{D}_{1;L,0}(z)}+\underbrace{\frac{1}{2}\Gamma\left(\frac{2z-1}{4}\right)\zeta_R\left(z-\frac{1}{2}\right)(2\pi r T)^{\frac{1}{2}-z}}_{\mathscr{D}_{1;L,1}(z) \left(m=0\;\text{term}\right)}\\&+\underbrace{\frac{1}{\sqrt{\pi}rT}\sum_{l=1}^{\infty}\sum_{m=1}^{\infty}
\left(\frac{l}{2mrT}\right)^{\frac{2z-3}{4}}K_{\frac{2z-3}{4}}\left(\frac{lm}{rT}\right)}_{\mathscr{D}_{1;L,2}(z)},\end{split}
\end{equation} \begin{equation}\label{eq6_2_5}
\begin{split}
\mathscr{D}_2(z)=& \underbrace{\frac{1}{4\sqrt{\pi}rT}\Gamma\left(\frac{ 2z+3}{4}\right)\zeta_R\left(z-\frac{1}{2}\right)}_{\mathscr{D}_{2;L,0}(z)}+\underbrace{\frac{1}{\sqrt{\pi}rT}\sum_{l=1}^{\infty}\sum_{m=1}^{\infty}\frac{1}{m^{\frac{2z-1}{4}}}
\left(\frac{l}{2 rT}\right)^{\frac{2z+3}{4}}K_{\frac{2z+3}{4}}\left(\frac{lm}{rT}\right)}_{\mathscr{D}_{2;L,2}(z)},
\end{split}
\end{equation}or\begin{equation}\label{eq6_2_2}
\begin{split}
\mathscr{D}_1(z)=&   \underbrace{\frac{1}{2}\Gamma\left(\frac{2z-1}{4}\right)\zeta_R\left(z-\frac{ 1}{2}\right) }_{\mathscr{D}_{1;T,0}(z) \left(l=0\;\text{  term}\right)}+ \underbrace{\frac{\sqrt{\pi}}{2}\Gamma\left(\frac{ 2z-3}{4}\right)\zeta_R\left(z-\frac{3}{2}\right)(2\pi rT)^{\frac{3}{2}-z}}_{\mathscr{D}_{1;T,1}(z)  }\\&+2\sqrt{\pi} \sum_{l=1}^{\infty}\sum_{m=1}^{\infty}
\left(\frac{m}{2   lrT}\right)^{\frac{2z-3}{4}}K_{\frac{2z-3}{4}}\left(4\pi^2 l m rT\right),\end{split}
\end{equation} \begin{equation}\label{eq6_2_6}
\begin{split}
\mathscr{D}_2(z)=& \underbrace{\frac{1}{2}\Gamma\left(\frac{2z+5}{4}\right)\zeta_R\left(z+\frac{1}{2}\right)}_{\mathscr{D}_{2;T,0}(z) \left(l=0\;\text{  term}\right)}+\underbrace{\frac{\sqrt{\pi}}{4}\Gamma\left(\frac{ 2z-1}{4}\right)\zeta_R\left(z-\frac{1}{2}\right)(2\pi rT)^{\frac{1}{2}-z}}_{\mathscr{D}_{2;T,1}(z)  }\\&+\sqrt{\pi} \sum_{l=1}^{\infty}\sum_{m=1}^{\infty}
\left(\frac{m}{2 lrT}\right)^{\frac{2z-1}{4}}K_{\frac{2z-1}{4}}\left(4\pi^2 l m rT\right)-
2 \pi^{\frac{5}{2}} \sum_{l=1}^{\infty}\sum_{m=1}^{\infty}
 \frac{m^{\frac{2z-1}{4}}}{(2  lrT)^{\frac{2z-5}{4}}} K_{\frac{2z-5}{4}}\left(4\pi^2 l m rT\right).
\end{split}
\end{equation}From these, one can deduce that $\mathscr{D}_1(z)$ only has simple poles at $z=5/2$ and $z=1/2$, whereas $\mathscr{D}_2(z)$ only has simple pole at $z=3/2$.

Now we can derive the asymptotic behaviors of the Casimir energy at small separation in different temperature regions. First we consider the leading order term.
Substituting \eqref{eq6_2_1} into \eqref{eq6_2_3}, we can obtain the low temperature leading behavior. In fact, the first term in \eqref{eq6_2_1}, $\mathscr{D}_{1;L,0}(z)$, is the term corresponding to the zero temperature Casimir energy. The residue at the largest pole at $z=5/2$ gives the zero temperature leading term:
\begin{equation*}
\begin{split}
E_{\text{Cas}}^{0,T=0}\sim &-\frac{ TL}{2\pi  r}\sum_{s=0}^{\infty}
 \text{Res}_{z=\frac{5}{2}}\left(\Gamma(z)(2\vep)^{-z}\frac{\mathscr{D}_{1;L,0}(z)}{\Gamma\left( \frac{2z+1}{4}\right)}\frac{1}{(s+1)^{z+\frac{3}{2}}}   \right)
=  -\frac{\pi^3L }{1920\sqrt{2}r^2\vep^{\frac{5}{2}}},
\end{split}
\end{equation*}which is exactly the leading term obtained by using proximity force approximation. Although   $\mathscr{D}_{1;L,0}(z)$ also has a pole at $z=3/2$, but this is canceled out by the pole at $z=3/2$ of the second term $\mathscr{D}_{1;L,1}(z)$, which is the term corresponding to   $m=0$.

Next, consider the contribution of the term $ \mathscr{D}_{1;L,2}(z)$ in \eqref{eq6_2_1} to the thermal correction of the Casimir energy. Since this term does not have any poles, we find that the leading term of the thermal correction is of order $\vep^0$, coming from the pole of $\Gamma(z)$ at $z=0$. This term would not contribute to the Casimir force since it is independent of $\vep$. Therefore, let us also consider the term of higher order in $\vep$ by taking the residue at $z=-1,-2,\ldots$. We have
\begin{equation}\label{eq6_2_9}
\begin{split}
\Delta_{T,2}E_{\text{Cas}}^{0 }\sim &-\frac{ TL}{2\pi  r}\sum_{j=0}^{\infty}\sum_{s=0}^{\infty}
 \text{Res}_{z= -j} \left(\Gamma(z)(2\vep)^{-z}\frac{\mathscr{D}_{1;L,2}(z)}{\Gamma\left( \frac{2z+1}{4}\right)}\frac{1}{(s+1)^{z+\frac{3}{2}}}   \right)\\
 =& -\sum_{j=0}^{\infty}\underbrace{\frac{(-1)^j}{j!}\frac{ 2^{j-1}\vep^j L}{\pi^{\frac{3}{2}}  r^2}\frac{\zeta_R\left(\frac{3}{2}-j\right)}{\Gamma\left(\frac{1-2j}{4}\right)}\sum_{l=1}^{\infty}\sum_{m=1}^{\infty}
\left(\frac{l}{2mrT}\right)^{-\frac{ 2j+3}{4}}K_{\frac{ 2j+3}{4}}\left(\frac{lm}{rT}\right)}_{\text{order $\vep^j$ term}}.
\end{split}
\end{equation}Notice that for the residue at $z=-j$, $j\leq 1$, the summation over $s$ is divergent, but we analytically continued it to obtain the value $\zeta_R(3/2-j)$.

The terms in \eqref{eq6_2_9} are exponentially small when $rT\ll 1$. But we cannot use this to conclude that the thermal correction is exponentially small when $rT\ll 1$. The Debye asymptotic expansion cannot be used to study the asymptotic behavior of the thermal correction in the low temperature region since all the terms in the asymptotic expansion would contribute, and summing over infinitely exponentially small terms would lead to polynomial order terms. Nevertheless, we can use \eqref{eq6_2_9} to study the leading behavior of the thermal correction in the medium temperature region where $a T\ll 1\ll rT$. Using the formula \eqref{eq6_1_4} again, we have
\begin{equation}\label{eq6_2_12}
\begin{split}
&\sum_{l=1}^{\infty}\sum_{m=1}^{\infty}
\left(\frac{l}{2mrT}\right)^{-\frac{ 2j+3}{4}}K_{\frac{ 2j+3}{4}}\left(\frac{lm}{rT}\right) \\=&\frac{1}{2}\sum_{l=1}^{\infty}\sum_{m=1}^{\infty}
\int_0^{\infty}t^{-\frac{2j+3}{4}-1}\exp\left(-\frac{1}{t}\left(\frac{l}{2rT}\right)^2-t m^2 \right)dt\\
\sim &\frac{1}{2}\sum_{l=1}^{\infty}\sum_{m=1}^{\infty}\frac{1}{2\pi i}\int_{c-i\infty}^{c+i\infty}\Gamma(w)\left(\frac{2rT}{l}\right)^{2w}\int_0^{\infty}t^{w-\frac{2j+3}{4}-1}\exp\left( -t m^2 \right)dtdz\\
=&\frac{1}{2}\frac{1}{2\pi i}\int_{c-i\infty}^{c+i\infty}\Gamma(w)(2rT)^{2w}\zeta_R(2w)\Gamma\left(w-\frac{2j+3}{4}\right)\zeta_R\left(2w -\frac{2j+3}{2}\right)dw\\
=&\frac{\sqrt{\pi}}{4} \Gamma\left(\frac{2j+5}{4}\right)\zeta_R\left(j+\frac{5}{2}\right)(2rT)^{j+\frac{5}{2}} -\frac{1}{4}
\Gamma\left(\frac{2j+3}{4}\right)\zeta_R\left(j+\frac{3}{2}\right)(2rT)^{j+\frac{3}{2}}   +O(T).
\end{split}
\end{equation}Notice that from the term $\mathscr{D}_{1;L,1}(z)$, we have
\begin{equation} \label{eq6_2_11}
\begin{split}
\Delta_{T,1}E_{\text{Cas}}^{0 }\sim &-\frac{ TL}{2\pi  r}\sum_{j=0}^{\infty}\sum_{s=0}^{\infty}
 \text{Res}_{z= -j} \left(\Gamma(z)(2\vep)^{-z}\frac{\mathscr{D}_{1;L,1}(z)}{\Gamma\left( \frac{2z+1}{4}\right)}\frac{1}{(s+1)^{z+\frac{3}{2}}}   \right)\\
 =& -\sum_{j=0}^{\infty} \frac{(-1)^j}{j!}\frac{ 2^{j-1}\vep^j TL}{\pi  r}\frac{\zeta_R\left(\frac{3}{2}-j\right)}{\Gamma\left(\frac{1-2j}{4}\right)}\times\frac{1}{2}\Gamma\left(\frac{-2j-1}{4}\right)\zeta_R\left(-j-\frac{1}{2}\right)(2\pi r T)^{\frac{1}{2}+j}\\
 =& -\sum_{j=0}^{\infty} \frac{(-1)^j}{j!}\frac{ 2^{j-1}\vep^j L}{\pi^{\frac{3}{2}}  r^2}\frac{\zeta_R\left(\frac{3}{2}-j\right)}{\Gamma\left(\frac{1-2j}{4}\right)}\times\frac{1}{4}
\Gamma\left(\frac{2j+3}{4}\right)\zeta_R\left(j+\frac{3}{2}\right)(2rT)^{j+\frac{3}{2}},
\end{split}
\end{equation}where we have used the identity
$$\Gamma\left(\frac{s}{2}\right)\zeta_R(s)=\pi^{s-\frac{1}{2}}\Gamma\left(\frac{1-s}{2}\right)\zeta_R(1-s).$$
This term cancel with the second term in the last line of \eqref{eq6_2_12}. Therefore, we find that in the medium temperature region, the leading asymptotic behavior of the thermal correction is given by
\begin{equation}\label{eq6_2_13}
\begin{split}
\Delta_{T}E_{\text{Cas}}^{0 }\sim &-\frac{L}{8\pi r^2}\sum_{j=0}^{\infty} \vep^j\frac{(-1)^j2^j}{j!} \frac{\zeta_R\left(\frac{3}{2}-j\right)}{\Gamma\left(\frac{1-2j}{4}\right)} \Gamma\left(\frac{2j+5}{4}\right)\zeta_R\left(j+\frac{5}{2}\right)(2rT)^{j+\frac{5}{2}}\\
=&-\frac{L\sqrt{r}T^{\frac{5}{2}}}{4\sqrt{2}\pi}\zeta_R\left(\frac{5}{2}\right)\zeta_R\left(\frac{3}{2}\right)-\vep Lr^{\frac{3}{2}}T^{\frac{7}{2}}  \frac{3 }{4\sqrt{2}\pi}\zeta_R\left(\frac{1}{2}\right)\zeta_R\left(\frac{7}{2}\right)+\vep^2 Lr^{\frac{5}{2}}T^{\frac{9}{2}}  \frac{15 }{8\sqrt{2}\pi}\zeta_R\left(-\frac{1}{2}\right)
\zeta_R\left(\frac{9}{2}\right)+\ldots,
\end{split}
\end{equation}coming from the first term of \eqref{eq6_2_12}. In fact, one can also obtain these terms by using the  term $\mathscr{D}_{1;T,1}$ in \eqref{eq6_2_2} and taking the residue of \eqref{eq6_2_3} at $j=0,-1,-2,\ldots$. The first term in \eqref{eq6_2_13} is the leading term of the thermal correction to the Casimir energy in the medium temperature region. It agrees with the result of proximity force approximation \eqref{eq6_2_15}. For the Casimir force, taking derivative of \eqref{eq6_2_13} with respect to $a$ gives
\begin{equation}\label{eq6_2_14}
\begin{split}
\Delta_{T}F_{\text{Cas}}^{0 }\sim &  Lr^{\frac{3}{2}}T^{\frac{7}{2}}  \frac{3 }{4\sqrt{2}\pi}\zeta_R\left(\frac{1}{2}\right)\zeta_R\left(\frac{7}{2}\right)-\vep  Lr^{\frac{3}{2}}T^{\frac{9}{2}}  \frac{15 }{4\sqrt{2}\pi}\zeta_R\left(-\frac{1}{2}\right)
\zeta_R\left(\frac{9}{2}\right)+\ldots.
\end{split}
\end{equation}Again, the leading (first) term agree with the result of proximity force approximation \eqref{eq6_2_16}. In fact, the second term which is equal to $0.1851\vep r^{\frac{3}{2}}T^{\frac{9}{2}}$ can also be predicted by proximity force approximation (see \cite{6}).

Now let us consider the high temperature region. In this case, the Casimir energy is dominated by the terms with $l=0$. Substitute the $l=0$ term $\mathscr{D}_{1;T,0}$ in \eqref{eq6_2_2} into \eqref{eq6_2_3}, and taking the residue at the largest pole of $\mathscr{D}_{1;T,0}$ at $z=3/2$, we find that the leading term of the Casimir energy in the high temperature region is given by
\begin{equation*}
\begin{split}
E_{\text{Cas}}^{0, \text{cl} }\sim &-\frac{ TL}{2\pi  r}\sum_{s=0}^{\infty}
 \text{Res}_{z=\frac{3}{2}}\left(\Gamma(z)(2\vep)^{-z}\frac{\mathscr{D}_{1;T,0}(z)}{\Gamma\left( \frac{2z+1}{4}\right)}\frac{1}{(s+1)^{z+\frac{3}{2}}}   \right)=-\frac{LT}{16\sqrt{2}r \vep^{\frac{3}{2}}}\zeta_R(3),
\end{split}
\end{equation*}agreeing with the result of proximity force approximation \eqref{eq6_2_17}. As in the case of $\mathscr{D}_{1;L,0}$, the pole of $\mathscr{D}_{1;T,0}$ at $z=1/2$ would actually cancel with the $m=0$ term which is not explicit in this case.

From these discussions, we find that the leading order terms of the Casimir energy \eqref{eq6_2_3} which is derived from the exact formula agree with the proximity force approximations in the zero temperature, medium temperature and high temperature regions. Let us now consider the first order approximation from the term \eqref{eq6_2_4}. For the zero temperature Casimir energy, substitute $\mathscr{D}_{1;L,0}(z+1)$ and $\mathscr{D}_{2;L,0}(z)$ into \eqref{eq6_2_4}, and taking the residue at $z=3/2$ give
\begin{equation}\label{eq6_2_22}
\begin{split}
E_{\text{Cas}}^{1,D,T=0}\sim &-\frac{ TL}{2\pi  r}
 \sum_{s=0}^{\infty}\text{Res}_{z=\frac{3}{2}}\Biggl\{\Gamma(z)(2\vep)^{-z} \Biggl( \frac{\mathscr{D}_{2;L,0}(z)}{\Gamma\left(\frac{2z+7}{4}\right)}\left[\frac{(2z+7)(2z-3)}{48} \frac{1}{(s+1)^{z+\frac{1}{2}}}
  +\frac{(2z+1)(2z+3)}{24} \frac{1}{(s+1)^{z+\frac{5}{2}}}\right]\\
&\hspace{4cm}+ \frac{\mathscr{D}_{1;L,0}(z+1)}{\Gamma\left(\frac{2z+3}{4}\right)}\left[-\frac{2z-3}{12} \frac{1}{(s+1)^{z+\frac{1}{2}}}
  -\frac{4z+3}{48} \frac{1}{(s+1)^{z+\frac{5}{2}}}\right]\Biggr)\Biggr\}\\
  =& -\frac{7\pi^3L }{69120\sqrt{2}r^2\vep^{\frac{3}{2}}}=-\frac{\pi^3L }{1920\sqrt{2}r^2\vep^{\frac{5}{2}}}\times \frac{7}{36}\vep
\end{split}
\end{equation}for the case of Dirichlet boundary conditions. For the case of Neumann boundary conditions, \eqref{eq6_2_18} gives
\begin{equation}\label{eq6_2_23}
\begin{split}
E_{\text{Cas}}^{1,N,T=0}\sim &E_{\text{Cas}}^{1,D,T=0} -\frac{ TL}{2\pi  r}\sum_{s=0}^{\infty}
 \text{Res}_{z=\frac{3}{2}}\Biggl\{ \Gamma(z)(2\vep)^{-z} \Biggl\{ \frac{\mathscr{D}_{2;L,0}(z)}{\Gamma\left(\frac{2z+7}{4}\right)}-\frac{\mathscr{D}_{1;L,0}(z+1)}{\Gamma\left(\frac{2z+3}{4}\right)}\Biggr\} \frac{1}{(s+1)^{z+\frac{1}{2}}}\Biggr\}\\
  =& -\frac{7\pi^3L }{69120\sqrt{2}r^2\vep^{\frac{3}{2}}}+  \frac{ \pi L }{144\sqrt{2}r^2\vep^{\frac{3}{2}}}=-\frac{\pi^3L }{1920\sqrt{2}r^2\vep^{\frac{5}{2}}}\times \left(\frac{7}{36}-\frac{40}{3\pi^2}\right)\vep.
\end{split}
\end{equation}\eqref{eq6_2_22} and \eqref{eq6_2_23} agree with the   first order correction for the zero temperature Casimir energies \eqref{eq6_2_20} and \eqref{eq6_2_21}  obtained by Bordag \cite{1}.

For the thermal correction, we have
\begin{equation*}
\begin{split}
\Delta_TE_{\text{Cas}}^{1,D }\sim &-\frac{ TL}{2\pi  r}\sum_{j=0}^{\infty}\sum_{s=0}^{\infty}
 \text{Res}_{z=-j}\Biggl\{\Gamma(z)(2\vep)^{-z} \Biggl( \frac{\mathscr{D}_{2;T,1}(z)}{\Gamma\left(\frac{2z+7}{4}\right)}\left[\frac{(2z+7)(2z-3)}{48}\frac{1}{(s+1)^{z+\frac{1}{2}}}
  +\frac{(2z+1)(2z+3)}{24} \frac{1}{(s+1)^{z+\frac{5}{2}}}\right]\\
&\hspace{4cm}+ \frac{\mathscr{D}_{1;T,1}(z+1)}{\Gamma\left(\frac{2z+3}{4}\right)}\left[-\frac{2z-3}{12} \frac{1}{(s+1)^{z+\frac{1}{2}}}
  -\frac{4z+3}{48} \frac{1}{(s+1)^{z+\frac{5}{2}}}\right]\Biggr)\Biggr\}\\
  =& -\frac{T^{\frac{3}{2}}L}{24\sqrt{2}r^{\frac{1}{2}}}\zeta_R\left(-\frac{1}{2}\right)\left(2\zeta_R\left(\frac{1}{2}\right)-\zeta_R\left(\frac{5}{2}\right)\right)
  +\vep\frac{\pi Lr^{\frac{1}{2}}T^{\frac{5}{2}}}{6\sqrt{2}}\zeta_R\left(-\frac{3}{2}\right)\left(10\zeta_R\left(-\frac{1}{2}\right)+\zeta_R\left(\frac{3}{2}\right)\right)+\ldots
\end{split}
\end{equation*}for the case of Dirichlet boundary conditions. For Neumann boundary conditions, \eqref{eq6_2_18} gives
\begin{equation*}
\begin{split}
\Delta_TE_{\text{Cas}}^{1,N }\sim &\Delta_TE_{\text{Cas}}^{1,D } -\frac{ TL}{2\pi  r}\sum_{j=0}^{\infty}\sum_{s=0}^{\infty}
 \text{Res}_{z=-j}\Biggl\{ \Gamma(z)(2\vep)^{-z} \Biggl\{ \frac{\mathscr{D}_{2;T,1}(z)}{\Gamma\left(\frac{2z+7}{4}\right)}-\frac{\mathscr{D}_{1;T,1}(z+1)}{\Gamma\left(\frac{2z+3}{4}\right)}\Biggr\} \frac{1}{(s+1)^{z+\frac{1}{2}}}\Biggr\}\\
 =& -\frac{T^{\frac{3}{2}}L}{24\sqrt{2}r^{\frac{1}{2}}}\zeta_R\left(-\frac{1}{2}\right)\left(18\zeta_R\left(\frac{1}{2}\right)-\zeta_R\left(\frac{5}{2}\right)\right)
  +\vep\frac{\pi Lr^{\frac{1}{2}}T^{\frac{5}{2}}}{6\sqrt{2}}\zeta_R\left(-\frac{3}{2}\right)\left(-6\zeta_R\left(-\frac{1}{2}\right)+\zeta_R\left(\frac{3}{2}\right)\right)+\ldots.
\end{split}
\end{equation*}These terms are of order $1/(rT)$ smaller than the leading term. For the Casimir force, we find that
\begin{equation*}
\begin{split}
\Delta_TF_{\text{Cas}}^{1,D }\sim &-\frac{\pi LT^{\frac{5}{2}}}{6\sqrt{2}r^{\frac{1}{2}}}\left(10\zeta_R\left(-\frac{1}{2}\right)+\zeta_R\left(\frac{3}{2}\right)\right)+\ldots=-0.1975Lr^{-\frac{1}{2}}T^{\frac{5}{2}}
+\ldots,
\end{split}
\end{equation*}and
\begin{equation*}
\begin{split}
\Delta_TF_{\text{Cas}}^{1,N }\sim &-\frac{\pi LT^{\frac{5}{2}}}{6\sqrt{2}r^{\frac{1}{2}}}\left(-6\zeta_R\left(-\frac{1}{2}\right)+\zeta_R\left(\frac{3}{2}\right)\right)
+\ldots=-1.4290Lr^{-\frac{1}{2}}T^{\frac{5}{2}}+\ldots.
\end{split}
\end{equation*}

For the high temperature (classical) term, substitute $\mathscr{D}_{1;T,0}(z+1)$ and $\mathscr{D}_{2;T,0}(z)$ into \eqref{eq6_2_4}, and taking the residue at $z=1/2$ give
\begin{equation}\label{eq6_2_26}
\begin{split}
E_{\text{Cas}}^{1,D,\text{cl}}\sim &-\frac{ LT}{2\pi  r}\sum_{s=0}^{\infty}
 \text{Res}_{z=\frac{1}{2}}\Biggl\{\Gamma(z)(2\vep)^{-z} \Biggl( \frac{\mathscr{D}_{2;T,0}(z)}{\Gamma\left(\frac{2z+7}{4}\right)}\left[\frac{(2z+7)(2z-3)}{48} \frac{1}{(s+1)^{z+\frac{1}{2}}}
  +\frac{(2z+1)(2z+3)}{24} \frac{1}{(s+1)^{z+\frac{5}{2}}}\right]\\
&\hspace{4cm}+ \frac{\mathscr{D}_{1;T,0}(z+1)}{\Gamma\left(\frac{2z+3}{4}\right)}\left[-\frac{2z-3}{12} \frac{1}{(s+1)^{z+\frac{1}{2}}}
  -\frac{4z+3}{48} \frac{1}{(s+1)^{z+\frac{5}{2}}}\right]\Biggr)\Biggr\}\\
  =&\underbrace{-\frac{LT}{768\sqrt{2\pi}r\vep^{\frac{1}{2}}}\sum_{s=0}^{\infty}\text{Res}_{z=\frac{1}{2}}\left\{\frac{\Gamma\left(\frac{2z+1}{4}\right)}{\Gamma\left(\frac{2z+7}{4}\right)}(2z-3)(2z-1)(2z+5)
  \zeta_R\left(z+\frac{1}{2}\right)\frac{1}{(s+1)^{z+\frac{1}{2}}}\right\}}_{\Lambda_1}-\frac{LT}{64\sqrt{2}r\vep^{\frac{1}{2}}}\zeta_R(3)\\
  =&-\frac{LT}{64\sqrt{2}r\vep^{\frac{1}{2}}}\zeta_R(3).
\end{split}
\end{equation}
For the term $\Lambda_1$, we have taken the residue at $z=1/2$ before taking the sum with respect to $s$, which give the value zero because of the factor $(2z-1)$. If we sum over $s$ first before taking the residue, then we have the term $\zeta_R(z+1/2)$ which has a pole at $z=1/2$. Using the fact that $\displaystyle \lim_{z\rightarrow 1/2}(2z-1)\zeta_R(z+1/2)=2$, we find that we  should add
$$\frac{LT}{32\sqrt{2}r\vep^{\frac{1}{2}}}$$ to $E_{\text{Cas}}^{1,D,\text{cl}}$. In the former case where we do not sum over $s$ before taking the residue, we find that in the high temperature region, the Casimir energy has asymptotic behavior
$$E_{\text{Cas}}^{D,\text{cl}}\sim  -\frac{LT}{16\sqrt{2}r \vep^{\frac{3}{2}}}\zeta_R(3)\left(1+\frac{\vep}{4}+\ldots\right);$$and hence the Casimir force has asymptotic behavior
\begin{equation}\label{eq6_3_1}F_{\text{Cas}}^{D,\text{cl}}\sim  -\frac{3LT}{32\sqrt{2}r \vep^{\frac{5}{2}}}\zeta_R(3)\left(1+\frac{\vep}{12}+\ldots\right).\end{equation}But we sum over $s$ first, then
\begin{equation}\label{eq6_3_2}F_{\text{Cas}}^{D,\text{cl}}\sim  -\frac{3LT}{32\sqrt{2}r \vep^{\frac{5}{2}}}\zeta_R(3)\left(1+\vep\left(\frac{1}{12}-\frac{1}{6\zeta_R(3)}\right)+\ldots\right).\end{equation}
The term $\displaystyle\frac{1}{12}-\frac{1}{6\zeta_R(3)}=-0.0533$ is negative. In \cite{6}, Weber and Gies used worldline numerical method to show that the first order correction has the same sign as the leading term. Therefore, the result \eqref{eq6_3_1} is more agreeable. In fact, one would also obtain \eqref{eq6_3_1} if one approximate the summation over $m$ by an integral, as in the work of Bordag \cite{1} for the zero temperature Casimir energy.

For the case of Neumann boundary conditions, using \eqref{eq6_2_18}, one   finds that
\begin{equation*}
\begin{split}
E_{\text{Cas}}^{1,N,\text{cl}}\sim &E_{\text{Cas}}^{1,D,\text{cl}}-\frac{ LT}{2\pi  r}\sum_{s=0}^{\infty}
 \text{Res}_{z=\frac{1}{2}}\Biggl\{ \Gamma(z)(2\vep)^{-z} \Biggl\{ \frac{\mathscr{D}_{2;T,0}(z)}{\Gamma\left(\frac{2z+7}{4}\right)}-\frac{\mathscr{D}_{1;T,0}(z+1)}{\Gamma\left(\frac{2z+3}{4}\right)}\Biggr\} \frac{1}{(s+1)^{z+\frac{1}{2}}}\Biggr\}\\
 =&E_{\text{Cas}}^{1,D,\text{cl}}+\frac{ LT}{8\pi  r}\sum_{s=0}^{\infty}
 \text{Res}_{z=\frac{1}{2}}\Biggl\{ \Gamma(z)(2\vep)^{-z} \frac{\Gamma\left(\frac{2z+1}{4}\right)}{\Gamma\left(\frac{2z+7}{4}\right)} \zeta_R\left(z+\frac{1}{2}\right)  \frac{1}{(s+1)^{z+\frac{1}{2}}}\Biggr\}\\
 =&E_{\text{Cas}}^{1,D,\text{cl}}+\frac{ LT}{8\sqrt{2}  r\vep^{\frac{1}{2}}}\sum_{s=0}^{\infty}\frac{1}{s+1},
\end{split}
\end{equation*}which is divergent. If we sum over $s$ first, then
\begin{equation*}\begin{split}
E_{\text{Cas}}^{1,N,\text{cl}}\sim &
 E_{\text{Cas}}^{1,D,\text{cl}}+\frac{ LT}{8\pi  r}
 \text{Res}_{z=\frac{1}{2}}\Biggl\{ \Gamma(z)(2\vep)^{-z} \frac{\Gamma\left(\frac{2z+1}{4}\right)}{\Gamma\left(\frac{2z+7}{4}\right)} \zeta_R\left(z+\frac{1}{2}\right)  \zeta_R\left(z+\frac{1}{2}\right)\Biggr\}\\
 =&-\frac{LT}{64\sqrt{2}r\vep^{\frac{1}{2}}}\zeta_R(3)+\frac{LT}{32\sqrt{2}r\vep^{\frac{1}{2}}}+\frac{ LT}{8\sqrt{2}  r\vep^{\frac{1}{2}}}
 \left(\psi\left(\frac{1}{2}\right)-\ln(2\vep)+\frac{1}{2}\psi\left(\frac{1}{2}\right)-\frac{1}{2}\psi(2)-2\psi(1)\right)\\
 =&-\frac{LT}{16\sqrt{2}r\vep^{\frac{1}{2}}}\left(2\ln\vep+\frac{\zeta_R(3)}{4}+8\ln 2-2\boldsymbol{C}+\frac{1}{2}\right),
\end{split}
\end{equation*}where $\boldsymbol{C}$ is the Euler constant. We see that there is a term of order $\vep\ln\vep$ for the first order correction. For the Casimir force, one finds that
\begin{equation*}
F_{\text{Cas}}^{N,\text{cl}}\sim  -\frac{3LT}{32\sqrt{2}r \vep^{\frac{5}{2}}}\zeta_R(3)\left(1+\vep\left(\frac{1}{12}+\frac{1}{ \zeta_R(3)}\left[\frac{2}{3}\ln\vep+\frac{8}{3}\ln 2-\frac{2}{3}\boldsymbol{C}-\frac{7}{6}\right]\right)+\ldots\right).\end{equation*}

From the discussion above, we see that there is some ambiguities regarding whether one should sum over $s$ first before taking the residue. The summation over $s$ appears because we have expanded the logarithm. The divergence in $s$ is possibly due to the fact that we have expanded each term in the expansion perturbatively, by treating $\vep$ and $n_i/m$ as small variables.   When we obtain a divergent sum in $s$, we need to question whether the perturbation is legitimate. For the leading order term of the Casimir force in the medium temperature region, we have found that correct result is obtained by summing over $s$ first, performing analytic continuation before taking the residue. However, for the first correction term to the Casimir energy in the high temperature region, this seems to lead to a result that does not agree with that obtained using numerical approach. The latter might also be special because we are facing the divergence series $\sum_{s=0}^{\infty}1/(s+1)$, which is formally $\zeta_R(1)$, infinity.

\section{Conclusion}

In this article, we have studied analytically the exact Casimir interaction between a cylinder and a plate at finite temperature for a scalar field with Dirichlet and Neumann boundary conditions. By taking the sum of the Dirichlet and Neumann Casimir interactions, one obtains the Casimir interaction for electromagnetic field with perfect conductor conditions. We   rederive the proximity force approximations to  the Casimir energy and the Casimir force in three temperature regions: the low temperature region where $aT\ll rT\ll 1$, the medium temperature region where $aT\ll 1\ll rT$ and the high temperature region where $1\ll aT\ll rT$. These were compared to the leading terms computed from the exact expression. It is shown that in the zero temperature, the medium temperature and the high temperature regions, the leading terms derived from the exact expression agree completely with the proximity force approximations. In fact, for the zero temperature case, the agreement between the proximity force approximation and the exact result has been shown in \cite{1}.  Using a similar perturbation developed in \cite{1}, we apply the inverse Mellin transform representation of the exponential function to derive the first analytic corrections to the proximity force approximations. In the zero temperature limit,   the results obtained in \cite{1} were recovered.

\begin{acknowledgments}
 This project is funded by the Ministry of Higher Education of Malaysia   under the FRGS grant FRGS/2/2010/SG/UNIM/02/2.
\end{acknowledgments}


\begin{thebibliography}{10}
\bibitem{48}
M. Bordag, G. L. Klimchitskaya, U. Mohideen and V. M. Mostepanenko, \emph{Advances in the Casimir effect}, Oxford University Press, Oxford, 2009.


\bibitem{34} S. K. Lamoreaux, \emph{Demonstration of the Casimir force in the 0.6 to 6 $\mu$m range}, Phys. Rev. Lett. \textbf{78}, 5 (1997).

\bibitem{35} U. Mohideen and A. Roy, \emph{Precision measurement of the Casimir force from 0.1 to 0.9 $\mu$m}, Phys. Rev. Lett. \textbf{81}, 4549 (1998).

\bibitem{36} Chen, F., G. L. Klimchitskaya, U. Mohideen, and V. M. Mostepanenko,  \emph{Theory confronts experiment in the Casimir force measurements:
Quantification of errors and precision}, Phys. Rev. A \textbf{69}, 022117 (2004).

\bibitem{37}
Decca, R. S., D. L$\acute{\text{o}}$pez, E. Fischbach, G. L. Klimchitskaya, D.
E. Krause, and V. M. Mostepanenko, \emph{Tests of new physics from precise measurements of the Casimir pressure
between two gold-coated plates}, Phys. Rev. D \textbf{75}, 077101 (2007).

\bibitem{38}
Decca, R. S., D. L$\acute{\text{o}}$pez, E. Fischbach, G. L. Klimchitskaya, D.
E. Krause, and V. M. Mostepanenko,  \emph{Novel constraints on light elementary particles
and extra-dimensional physics from the Casimir effect}, Eur. Phys. J. C
\textbf{51}, 963 (2007).

\bibitem{39} G. L. Klimchitskaya, U. Mohideen and V. M. Mostepanenko,  \emph{The Casimir force between real materials: Experiment and theory}, Rev. Mod. Phys. \textbf{81}, 1827 (2009).


\bibitem{40} G. Bressi, G. Carugno, R. Onofrio and G. Ruoso, \emph{Measurement of the Casimir force between parallel metallic surfaces}, Phys. Rev. Lett. \textbf{88}, 041804 (2002).





\bibitem{42} Q. Wei, D. A. R. Dalvit, F. C. Lombardo, F. D. Mazzitelli and R. Onofrio, \emph{Results from electrostatic calibrations for measuring the Casimir force
in the cylinder-plane geometry}, Phys. Rev. A \textbf{81}, 052115 (2010).

\bibitem{43} M. Schaden and L. Spruch, \emph{Infinity-free semiclassical evaluation of Casimir effects}, Phys. Rev. A \textbf{58}, 935 (1998).

\bibitem{44} M. Schaden and L. Spruch, \emph{Focusing virtual photons: Casimir energies for some pairs of conductors}, Phys. Rev. Lett. \textbf{84}, 459 (2000).

\bibitem{45} R. L. Jaffe and A. Scardicchio, \emph{Casimir effect and geometric optics}, Phys. Rev. Lett. \textbf{92}, 070402 (2004).

\bibitem{46} A. Scardicchio and R. L. Jaffe, \emph{Casimir effects: An optical approach I. Foundations and examples}, Nucl. Phys. B \textbf{704}, 552 (2005).

\bibitem{47} A. Scardicchio and R. L. Jaffe, \emph{Casimir effects: An optical approach II. Local observables and thermal corrections}, Nucl. Phys. B \textbf{743}, 249 (2006).

\bibitem{14} H. Gies, K. Langfeld, and L. Moyaerts, \emph{Casimir effect on the worldline}, J. High Energy Phys. \textbf{0306}, 018 (2003).
\bibitem{15} H. Gies  and K. Klingm\"uller, \emph{Casimir effect for curved geometries: Proximity-force-approximation validity limits}, Phys. Rev. Lett. \textbf{96}, 220401 (2006).
\bibitem{16} H. Gies  and K. Klingm\"uller,  \emph{Casimir edge effects}, Phys. Rev. Lett. \textbf{97}, 220405 (2006).
\bibitem{17} H. Gies  and K. Klingm\"uller, \emph{Worldline algorithms for Casimir configurations}, Phys. Rev. D \textbf{74}, 045002 (2006).

\bibitem{1} M. Bordag, \emph{Casimir effect for a sphere and a cylinder in front of a plane and corrections to the proximity force theorem}, Phys. Rev. D \textbf{73}, 125018 (2006).
\bibitem{27} A. Bulgac, P. Magierski and A. Wirzba, \emph{Scalar Casimir effect between Dirichlet spheres or a plate and a sphere}, Phys. Rev. D \textbf{73}, 025007 (2006).

\bibitem{3} T. Emig, R. L. Jaffe, M. Kadar and A. Scardicchio, \emph{Casimir interaction between a plate and a cylinder}, Phys. Rev. Lett. \textbf{96}, 080403 (2006).
\bibitem{9} T. Emig, N. Graham, R. L. Jaffe and M. Kardar, \emph{Casimir forces between arbitrary compact objects}, Phys. Rev. Lett. \textbf{99}, 170403 (2007).

\bibitem{10} T. Emig, N. Graham, R. L. Jaffe and M. Kardar, \emph{Casimir forces between compact objects: The scalar case}, Phys. Rev. D \textbf{77}, 025005 (2008).
\bibitem{11} T. Emig and R. L. Jaffe, \emph{Casimir forces between arbitrary compact objects}, J. Phys. A \textbf{41}, 164001 (2008).
\bibitem{12} O. Kenneth and I. Klich, \emph{Casimir forces in a T-operator approach}, Phys. Rev. B \textbf{78}, 014103 (2008).
\bibitem{32} K. A. Milton and J. Wagner, \emph{Multiple scattering methods in Casimir calculations}, J. Phys. A \textbf{41}, 155402 (2008).

\bibitem{13} S. J. Rahi, T. Emig, N. Graham, R. L. Jaffe, and M. Kardar, \emph{Scattering theory approach to electrodynamic Casimir forces}, Phys. Rev. D \textbf{80}, 085021 (2009).

\bibitem{21} D. A. R. Dalvit, F. C. Lombardo, F. D. Mazzitelli and R. Onofrio, \emph{Exact Casimir interaction between eccentric cylinders}, Phys. Rev. A \textbf{74}, 020101(R) (2006).
\bibitem{2} F. D. Mazzitelli, D. A. R. Dalvit and F. C. Lombardo, \emph{Exact zero-point interaction energy between cylinders}, New. J. Phys. \textbf{8}, 240 (2006).

\bibitem{19} G. L. Klimchitskaya, U. Mohideen and V. M. Mostepanenko, \emph{Casimir and Van der Waals forces between two plates or a sphere (lens) above a plate made of real materials}, Phys. Rev. A \textbf{61}, 062107 (2000).

\bibitem{23} M. Bordag and V. Nikolaev, \emph{Casimir force for a sphere in front of a plane beyond proximity force approximation}, J. Phys. A \textbf{41}, 164002 (2008).

\bibitem{20} A. Canaguier-Durand, P. A. Maia Neto, I. Cavero-Pelaez, A. Lambrecht, and S. Reynaud, \emph{Casimir interaction between plane and spherical metallic surfaces}, Phys. Rev. Lett. \textbf{102}, 230404 (2009).



\bibitem{29} A. Canaguier-Durand, A. G$\acute{\text{e}}$rardin, R. Gu$\acute{\text{e}}$rout, P. A. Maia Neto, V. V. Nesvizhevsky, A. Yu. Voronin, A. Lambrecht, and S. Reynaud, \emph{Casimir interaction between a dielectric nanosphere and a metallic plane}, Phys. Rev. A \textbf{83}, 032508 (2011).

\bibitem{31} M Bordag and V Nikolaev, \emph{First analytic correction beyond the proximity force approximation in the Casimir effect
for the electromagnetic field in sphere-plane geometry}, Phys. Rev. D \textbf{81}, 065011 (2010).



\bibitem{22} F. C. Lombardo, F. D. Mazzitelli and P. I. Villar, \emph{Numerical evaluation of the Casimir interaction between cylinders}, Phys. Rev. D \textbf{78}, 085009 (2008).


\bibitem{4} S. J. Rahi, T. Emig, R. L. Jaffe and M. Kadar, \emph{Casimir forces between cylinders and plates}, Phys. Rev. A \textbf{78}, 012104 (2008).


\bibitem{30} M Bordag and V Nikolaev, \emph{The vacuum energy for two cylinders with one increasing in size}, J. Phys. A  \textbf{42},  415203 (2009).


\bibitem{33} F. C. Lombardo, F. D. Mazzitelli and P. I. Villar, \emph{Exploring the quantum vacuum with cylinders}, J. Phys. A \textbf{41}, 164009 (2008).




\bibitem{24} S. Zaheer, S. J. Rahi, T. Emig and R. L. Jaffe, \emph{Casimir interactions of an object inside a spherical metal shell}, Phys. Rev. A \textbf{81}, 030502 (2010).
\bibitem{25} S. Zaheer, S. J. Rahi, T. Emig and R. L. Jaffe, \emph{Casimir potential of a compact object enclosed by a spherical cavity}, Phys. Rev. A \textbf{82}, 052507 (2010).




\bibitem{7}  A. Weber and H. Gies, \emph{Nonmonotonic thermal Casimir force from geometry-temperature interplay}, Phys. Rev. Lett. \textbf{105}, 040403 (2010).
\bibitem{6} A. Weber and H. Gies, \emph{Geothermal Casimir phenomena for the sphere-plate and cylinder-plate configurations}, Phys. Rev. D \textbf{82}, 125019 (2010).

\bibitem{49}  A. Canaguier-Durand, P. A. Maia Neto,  A. Lambrecht, and S. Reynaud, \emph{Thermal Casimir effect in the plane-sphere geometry
}, Phys. Rev. Lett. \textbf{104}, 040403 (2010).

\bibitem{28}  A. Canaguier-Durand, P. A. Maia Neto,  A. Lambrecht, and S. Reynaud, \emph{Thermal Casimir effect for Drude metals in the plane-sphere geometry}, Phys. Rev. A \textbf{82}, 012511 (2010).

\bibitem{5} M. Bordag and I. Pirozhenko, \emph{Vacuum energy between a sphere and a plane at finite temperature}, Phys. Rev. D \textbf{81}, 085023 (2010).

\bibitem{26} M. Bordag, I. G. Pirozhenko, \emph{Casimir entropy for a ball in front of a plane}, Phys. Rev. D \textbf{82}, 125016 (2010).

\bibitem{50} D. A. R. Dalvit, F. C. Lombardo, F. D. Mazzitelli and R. Onofrio, \emph{Casimir force between eccentric cylinders}, Eur. Phys. Lett. \textbf{67}, 517 (2004).

\bibitem{41} M. Brown-Hayes, D. A. R. Dalvit, F. D. Mazzitelli, W. J. Kim and R. Onofrio, \emph{Towards a precision measurement of the Casimir force in a cylinder-plane geometry}, Phys. Rev. A \textbf{72}, 052102 (2005).

\bibitem{18} G. L. Klimchitskaya and C. Romero, \emph{Strengthening constraints on Yukawa-type corrections to Newtonian gravity from measuring the Casimir force between a cylinder and a plate}, Phys. Rev. D \textbf{82}, 115005 (2010).



\bibitem{8} S. C. Lim and L. P. Teo, \emph{Finite-temperature Casimir effect in piston geometry and its classical limit}, Eur. Phys. J. C. \textbf{60}, 323 (2009).

\end{thebibliography}
\end{document}